\documentclass[letterpaper,english,prl,twocolumn,groupedaddress,reprint]{revtex4}
\usepackage[T1]{fontenc}
\usepackage[latin9]{inputenc}
\usepackage{mathrsfs}
\usepackage{amsmath}
\usepackage{amssymb}
\usepackage{graphicx}
\usepackage{braket}
\usepackage{xcolor}
\usepackage{comment}
\makeatletter

\@ifundefined{textcolor}{}
{%
 \definecolor{BLACK}{gray}{0}
 \definecolor{WHITE}{gray}{1}
 \definecolor{RED}{rgb}{1,0,0}
 \definecolor{GREEN}{rgb}{0,1,0}
 \definecolor{BLUE}{rgb}{0,0,1}
 \definecolor{CYAN}{cmyk}{1,0,0,0}
 \definecolor{MAGENTA}{cmyk}{0,1,0,0}
 \definecolor{YELLOW}{cmyk}{0,0,1,0}
}

\usepackage[unicode=true,
 bookmarks=false,
 breaklinks=false,pdfborder={0 0 1},backref=false,colorlinks=true]
 {hyperref}
\hypersetup{
 linkcolor=blue,citecolor=blue,urlcolor=blue,linkcolor=blue,citecolor=blue,urlcolor=blue}
\usepackage{dsfont}

\AtBeginDocument{
  
}

\makeatother

\usepackage{babel}
\begin{document}

\title{Quantum Simulation and Optimization in Hot Quantum Networks}

\author{M.J.A. Schuetz,$^{1,*}$ B. Vermersch,$^{2,3,*}$ G. Kirchmair,$^{2,3}$ L.M.K. Vandersypen,$^{4}$ J.I. Cirac,$^{5}$ M.D. Lukin,$^{1}$  and P. Zoller$^{2,3}$}

\affiliation{$^{1}$Physics Department, Harvard University, Cambridge, MA 02318,USA}
\affiliation{$^{2}$Center for Quantum Physics, and Institute for Experimental Physics, University of Innsbruck, A-6020 Innsbruck, Austria}
\affiliation{$^{3}$Institute for Quantum Optics and Quantum Information of the Austrian Academy of Sciences, A-6020 Innsbruck, Austria}
\affiliation{$^{4}$QuTech and Kavli Institute of NanoScience, TU Delft, 2600 GA Delft, The Netherlands}
\affiliation{$^{5}$Max-Planck-Institut f\"ur Quantenoptik, Hans-Kopfermann-Str. 1, 85748 Garching, Germany}

\date{\today}
\begin{abstract}
We propose and analyze a setup based on (solid-state) qubits coupled to a common multi-mode transmission line, which allows for coherent spin-spin interactions over macroscopic on-chip distances, without any ground-state cooling requirements for the data bus. 
Our approach allows for the realization of fast deterministic quantum gates between distant qubits, 
the simulation of quantum spin models with engineered (long-range) interactions, 
and provides a flexible architecture for the implementation of quantum approximate optimization algorithms. 
\end{abstract}

\maketitle

\begin{figure}
\includegraphics[width=0.95 \columnwidth]{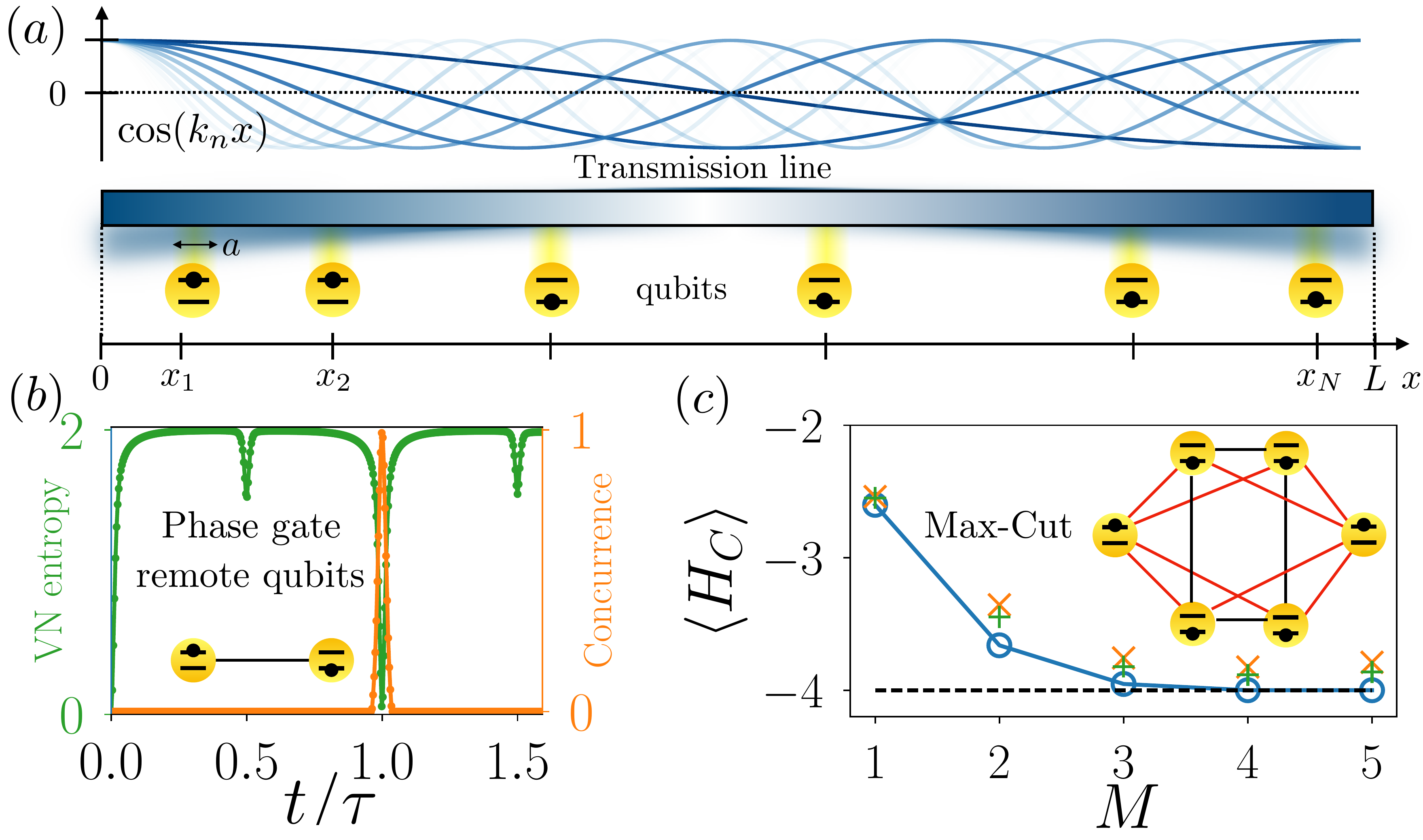}
\caption{{\it Hot Quantum Network.} (a) Schematic illustration of $N$ qubits coupled to a transmission line of length $L$. 
(b) Dynamic evolution of two qubits, as exemplified for the von Neumann (VN) entropy (left axis) and the concurrence (right axis) of the two-qubit density matrix, with $a=0.03L$. 
At the round trip time $t=\tau$, the qubits fully decouple from the waveguide and form a maximally entangled state, even though the transmission line is far away from the ground state (here, $k_B T=\omega_1$).
(c) Quantum approximate optimization algorithm (QAOA) solving Max-Cut with $N=6$ qubits and a $4$-regular graph (inset), in the presence of decoherence (ideal case: blue, dephasing with rate $\gamma_\phi/J_{\max}=0.003$: orange, rethermalization with rate $\kappa/|\Delta|=0.004$: green), and at finite temperature $k_B T=\omega_1$. 
Further details are given in the text.
\label{fig:setup}}
\end{figure}

\textit{Introduction.---}One of the leading approaches for scaling up quantum information systems involves  a modular architecture that makes use of a combination of short and long-distant interactions between the qubits \cite{monroe16, vandersypen17}.  
In particular, long-distant interactions can be implemented via a quantum bus which can effectively distribute quantum information between remote qubits, 
as shown in the context of of trapped ions~\cite{Poyatos1998,Molmer1999, milburn99,Ripoll2005, lemmer13}, 
solid state systems~\cite{schuetz17,Royer2017}, electromechanical resonators~\cite{rabl10}, as well as 
circuit QED architectures \cite{scarlino18, woerkom18, Gambetta2017, wendin17,hanson07, zwanenburg13}.
In this Letter, we provide a unified  theoretical framework for robust distribution of quantum information via a quantum bus that operates at finite temperature \cite{temperature}, 
fully accounts for the multi-mode structure of the data bus, 
and does not require the qubits to be identical.
Our approach [c.f Fig.~\ref{fig:setup}(a)]  results in an architecture where fully programmable interactions between qubits can be realized in a fast and deterministic way, without any ground-state cooling requirements for the data bus,
thereby setting the stage for various applications in the context of quantum information processing~\cite{Northup2014} in a hot quantum network, different from quantum state transfer discussed previously~\cite{cirac97, Vermersch2017,Xiang2017}. 
As illustrated in Fig.~\ref{fig:setup}(b), and discussed in detail below, one can use our scheme to deterministically implement (hot) quantum gates between two qubits. 
Moreover, we present a recipe to generate a targeted and scalable evolution for a large set of $N$ qubits coupled via a single transmission line,
thereby providing a 
natural architecture for the implementation of
quantum algorithms, such as quantum annealing~\cite{Das2008} or 
the quantum approximate optimization algorithm (QAOA)~\cite{farhi14, farhi16, otterback17}, designed to find approximate solutions to hard, combinatorial search problems. 

\textit{The model.---}We consider a set of qubits $i=1,2,\dots,N$ with corresponding transition frequencies $\omega_{i}$ (typically in the microwave regime) that are coupled to a 
(multi-mode) transmission line of length $L$; compare Fig.~\ref{fig:setup} for a schematic illustration. 
The transmission line is described in terms of photonic modes $a_n$ with wave-vectors $k_n=n\pi/L$, 
with a \textit{linear} spectrum $\omega_n=k_nc=n\omega_1$, where $\omega_1=\pi c/L$ is the frequency of the fundamental mode $n=1$ and $c$ is the (effective) speed of light. 
As opposed to transversal (Jaynes-Cummings-like) spin-resonator coupling,
here we focus on \textit{longitudinal} coupling as could be realized (for example) with superconducting qubits~\cite{Kerman2013,Billangeon2015,Didier2015,Richer2016,Royer2017}
or quantum dot based qubits \cite{childress04, harvey18, schuetz17, Royer2017, jin12, beaudoin16, russ17}.
The Hamiltonian under consideration then reads ($\hbar=1$)
\begin{equation}
H=\sum_{i=1}^N\frac{\omega_{i}}{2}\sigma_{i}^{z}+\sum_{n=1}^\infty \omega_n a_{n}^{\dagger}a_{n}+\sum_{i,n}  g_{i,n} \sigma_i^z \left(a_{n}+a_{n}^{\dagger}\right),\label{eq:model}
\end{equation}
with the Pauli matrices $\vec{\sigma}_{i}$ describing the qubits and $g_{i,n}$ the coupling strength between qubit $i$ and mode $n$.
We show below that for specific times $t$, which are integer multiples of the round-trip time $t\propto\tau\equiv 2L/c$, the dynamics of the qubits and \textit{all} photons fully decouple, while giving rise to an effective interaction between the qubits.

\textit{Analytical solution of time evolution.---}With the help of the spin-dependent, multi-mode displacement transformation
$U^{\dagger}_{\mathrm{pol}}=\exp[\sum_{n,i}\frac{g_{i,n}}{\omega_{n}}\sigma_{i}^{z}\left(a_{n}^{\dagger}-a_{n}\right)]$, 
in our model the spin dynamics can be decoupled from the resonator dynamics (in the polaron frame), and we find $H=U_{\mathrm{pol}} \tilde H U_{\mathrm{pol}} ^\dagger$, 
where
\begin{equation}
\tilde H=\sum_{i}\frac{\omega_{i}}{2}\sigma_{i}^{z}+\sum_{n}\omega_{n}a_{n}^{\dagger}a_{n}+\sum_{i<j} J_{ij}\sigma_{i}^{z}\sigma_{j}^{z}, 
\end{equation}
with the effective spin-spin interaction
\begin{equation}
J_{ij}=-2\sum_{n}\frac{g_{i,n}g_{j,n}}{\omega_{n}}. \label{eq:Jij_n_main}
\end{equation}
Therefore, the corresponding time-evolution in the lab frame reads $e^{-iHt}  =  U_{\mathrm{pol}}e^{-i \tilde{H}t} U_{\mathrm{pol}}^\dagger$. 
Consider now the evolution at stroboscopic times $t_p=p \tau$ ($p$ positive integer), corresponding to multiples of the round trip time $\tau$. 
In this case,  the \textit{synchronization} of the modes 
$\exp\left[-it_{p}\sum_{n}\omega_{n}a_{n}^{\dagger}a_{n}\right]=\exp\left[-2\pi i\sum_{n} n p a_{n}^{\dagger}a_{n}\right]=\mathds1$ 
implies that the full evolution
\textit{in the lab frame} reduces exactly to $U_{\mathrm{lab}}(t_{p}) = \exp[-iH t_{p}]$,
\begin{equation}
U_{\mathrm{lab}}(t_{p})=e^{-it_{p}\sum_{i}(\omega_{i}/2)\sigma_{i}^{z}}e^{-i t_{p}\sum_{i<j}J_{ij}\sigma_{i}^{z}\sigma_{j}^{z}}.\label{eq:unitary-stroboscopic}
\end{equation}
Accordingly, for certain times the qubits fully disentangle from the (thermally populated) resonator modes, thereby providing a qubit gate that is insensitive to the state of the resonator, 
while imposing no conditions on the qubit frequencies $\omega_{i}$. 
For specific times, the time evolution in the polaron and the laboratory frame coincide and fully decouple from the photon modes, allowing for the realization of a thermally robust gate, without any need of cooling the transmission line to the vacuum \cite{schuetz17}.
Moreover, our approach can be straightforwardly combined with standard spin-echo techniques in order to cancel out efficiently low-frequency noise: 
By synchronizing fast global $\pi$ rotations with the stroboscopic times $t_{p}$, one can enhance the qubit's coherence times from the time-ensemble-averaged dephasing time $T_{2}^{\star}$ to the prolonged timescale $T_{2}$. 

\textit{Frequency cutoff.---}In principle, the spin-spin coupling strength $J_{ij}$ as defined in Eq.~\eqref{eq:Jij_n_main} involves \textit{all} modes $n=1,2,\dots$, 
naively leading to unphysical divergencies, as discussed in the context of transversal qubit-resonator coupling in Refs.~\cite{filipp11, houck08}. 
In any physical implementation, however, there is a microscopic lengthscale $a$ that naturally introduces a frequency cutoff. 
Specifically, we take the coupling parameters $g_{i,n}$ as $g_{i,n}=g_i \sqrt{n} \int_{0}^L \cos (k_n x) f(x-x_i) dx$, to account for the fact that the qubits couple to the local 
voltage, where $f(x-x_i)$ accounts for the microscopic spatial extension of the qubit-transmission line coupling (cf. \cite{SM} for details);
the factor $\sim \sqrt{n}$ derives from the scaling of the rms zero-point voltage fluctuations with the mode index $n$, which also implies $g_i\propto  L^{-1}$.
In the examples below, we will consider for simplicity a box function 
$f(x)=\delta_{x>0}\delta_{x<a}/a$,
leading to 
$g_{i,n} = g_i \sqrt{n} \left( \sin\left[k_n(x_i+a) \right] -  \sin\left[k_n x_i \right] \right)/(k_{n}a)$.
Note that if the microscopic lengthscale $a$ is set to zero, yielding the (point-like) standard result $g_{i,n} = g_{i} \sqrt{n} \cos(k_{n}x_{i})$ \cite{sundaresan15}, the summation over $n$ in Eq.~\eqref{eq:Jij_n_main} does not converge.
Instead for a finite $a$, and for $|x_i-x_j| > a$ the effective spin-spin interaction Eq.~\eqref{eq:Jij_n_main} simplifies to $J_{ij}=g_{i}g_{j}/\omega_{1}$ (c.f. \cite{SM}).
Accordingly, within this exemplary model, the effective coupling $J_{ij}$ does not depend on the microscopic lengthscale $a$, nor the position of the qubits $x_i$, and scales as $L^{-1}$, 
as the rate at which interactions between qubits are generated is limited by the propagation time $\tau$ ($\propto L$) of light through the waveguide.

\begin{figure}[t]
\includegraphics[width=\columnwidth]{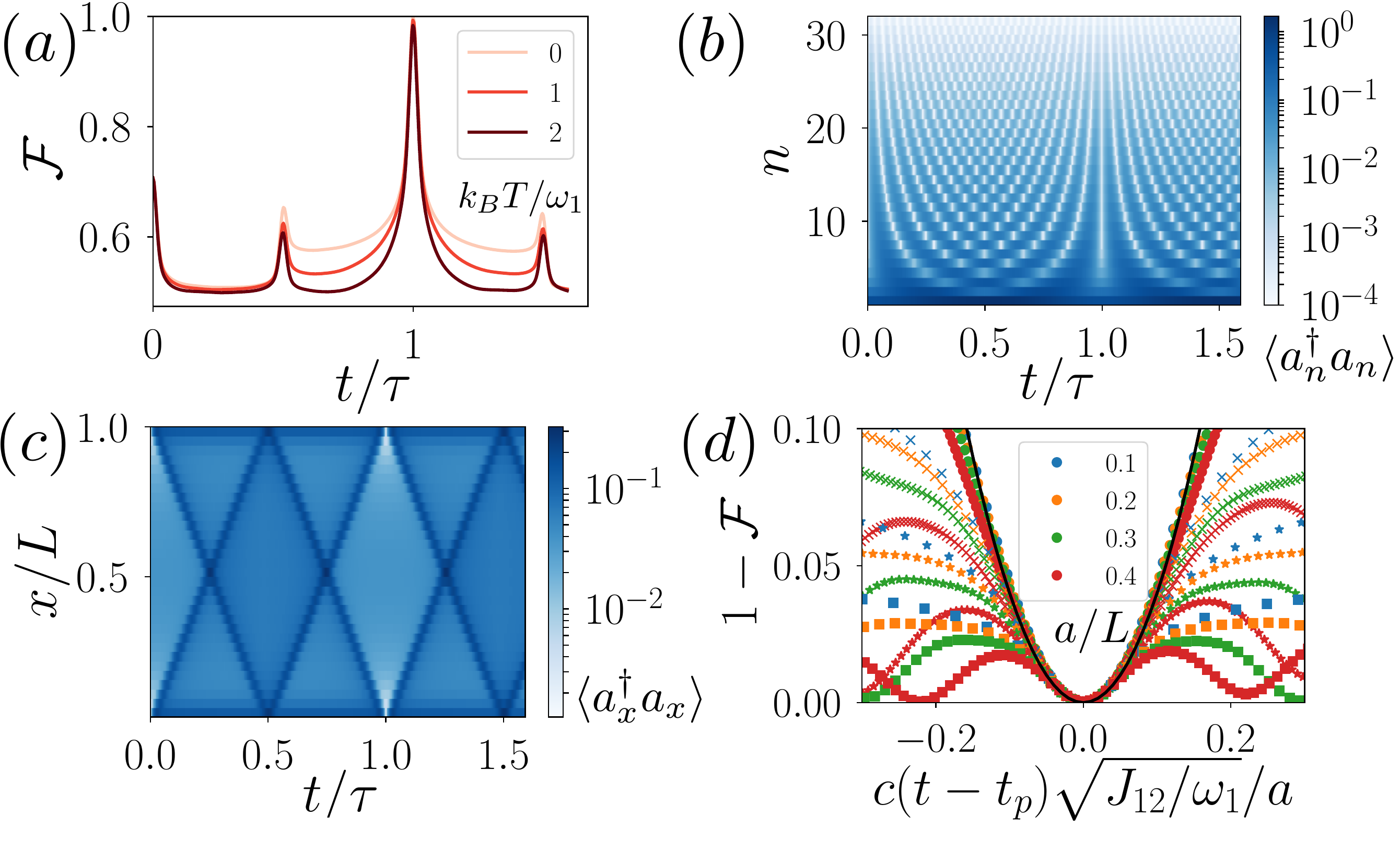}
\caption{{\it Hot phase gate between two distant qubits.} (a)-(b) Fidelity as a function of time $\tau$ (a) for $a=0.03L$ and different transmission line temperatures $0\le k_BT\le 2 \omega_1$. 
(b) Mode and (c) real space occupation as a function of the transmission line for $a=0.03L$ and $k_BT=\omega_1$, with $\sim 30$ modes.
(d) Error $1-\mathcal{F}$ around the gate time $t_p$ for $T=0$ and different values of the cutoff (legend) and number of cycles $p=1,4,8,16$ (circles, crosses, stars, squares). 
The black solid line refers to $4(c/a)^2 J_{12}/\omega_{1} \Delta t^2$. 
\label{fig:MPS}}
\end{figure}

\textit{Applications.---}In what follows, we discuss three applications of our scheme, with a gradual increase in complexity, 
namely (i) a \textit{hot} two-qubit phase gate, 
(ii) the engineering of spin models, and 
(iii) the implementation of QAOA in the presence of decoherence and finite temperature.  
To this end, we consider the possibility to potentially boost and fine-tune the effective spin-spin interactions $J_{ij}$ by parametrically modulating the longitudinal spin-resonator coupling, as could be realized with both superconducting qubits \cite{Royer2017} or quantum dot based qubits \cite{harvey18}; cf. \cite{SM} for further details.

\textit{Hot phase gate.---}As a first illustration of our scheme, we consider the realization of a phase-gate between two remote qubits $N=2$, placed at each edge of the transmission line ($x_1=0$, $x_2=L-a$).
Our initial state
$\rho_0=\ket{\Psi_0}\bra{\Psi_0}\otimes_n \rho_n$  
consists of a pure initial qubit state with $\ket{\Psi_0}=\otimes_j (\ket{0}+i\ket{1})_{j}/\sqrt{2}$ 
and a thermal state of the waveguide with $\rho_n=\exp(-\frac{a_n^\dagger a_n \omega_n}{k_B T})\left[1-\exp(-\frac{\omega_n}{k_B T})\right]$, 
and we use Matrix-Product-States (MPS) techniques~\cite{Peropadre2013} to show numerically how the hot quantum network generates the desired evolution Eq.~\eqref{eq:unitary-stroboscopic}.
We fix $g_i=\omega_{1}/\sqrt{8}$ which (under ideal circumstances) leads to a maximally entangled pure state \mbox{$\ket{\Psi(t_1)}=\exp(-i \frac{\pi}{4} \sigma_1^z \sigma_2^z)\ket{\Psi_0}$} after one round trip time $t_1=\tau$ (generalizations thereof are provided in \cite{SM}). 
In Fig.~\ref{fig:setup}(b), we show the von-Neumann entropy $\mathcal{E}$ and the concurrence $\mathcal{C}$ of the two-qubit density matrix $\rho_{1,2}$, showing the realization of the gate at $t=t_1$, in the presence of thermal occupation of the waveguide. 
The corresponding fidelity $\mathcal{F}$ defined as overlap of $\rho_{1,2}$ with respect to the ideal state $\ket{\Psi(t_1)}\bra{\Psi(t_1)}$ is shown in Fig.~\ref{fig:MPS}(a). 
In panels (b) and (c) both the mode occupation $\langle a^\dag_n a_n \rangle$ and the real space occupation $\langle a^\dagger_x a_x \rangle$ are displayed, with $a_x$, $0<x<L$, referring to the discrete sine transform of $a_n$.  
At the round trip time $t=\tau$, the waveguide returns to its initial thermal state, as expected. 
In panel (d), we study the scaling of timing errors by showing the evolution of the error $1-\mathcal{F}$ around $t \approx t_p$. In the limit of small errors, the numerical results are well approximated by 
$1-\mathcal{F} \approx 4(c/a)^2 J_{12}/\omega_{1} \Delta t^2$ (black line), with $\Delta t = t-t_p$. 
Accordingly, the timing error is sensitive to the cutoff $a$ (as it controls the frequency scale of the couplings), and scales linearly with the effective spin-spin interaction $J_{12}$, as slower dynamics are less vulnerable to timing inaccuracies $\sim \Delta t$; for further details, in particular related to the influence of temperature on timing errors, and effects due to nonlinear dispersion relations $\omega_n$, cf. \cite{SM}.

\begin{figure}
\includegraphics[width=\columnwidth]{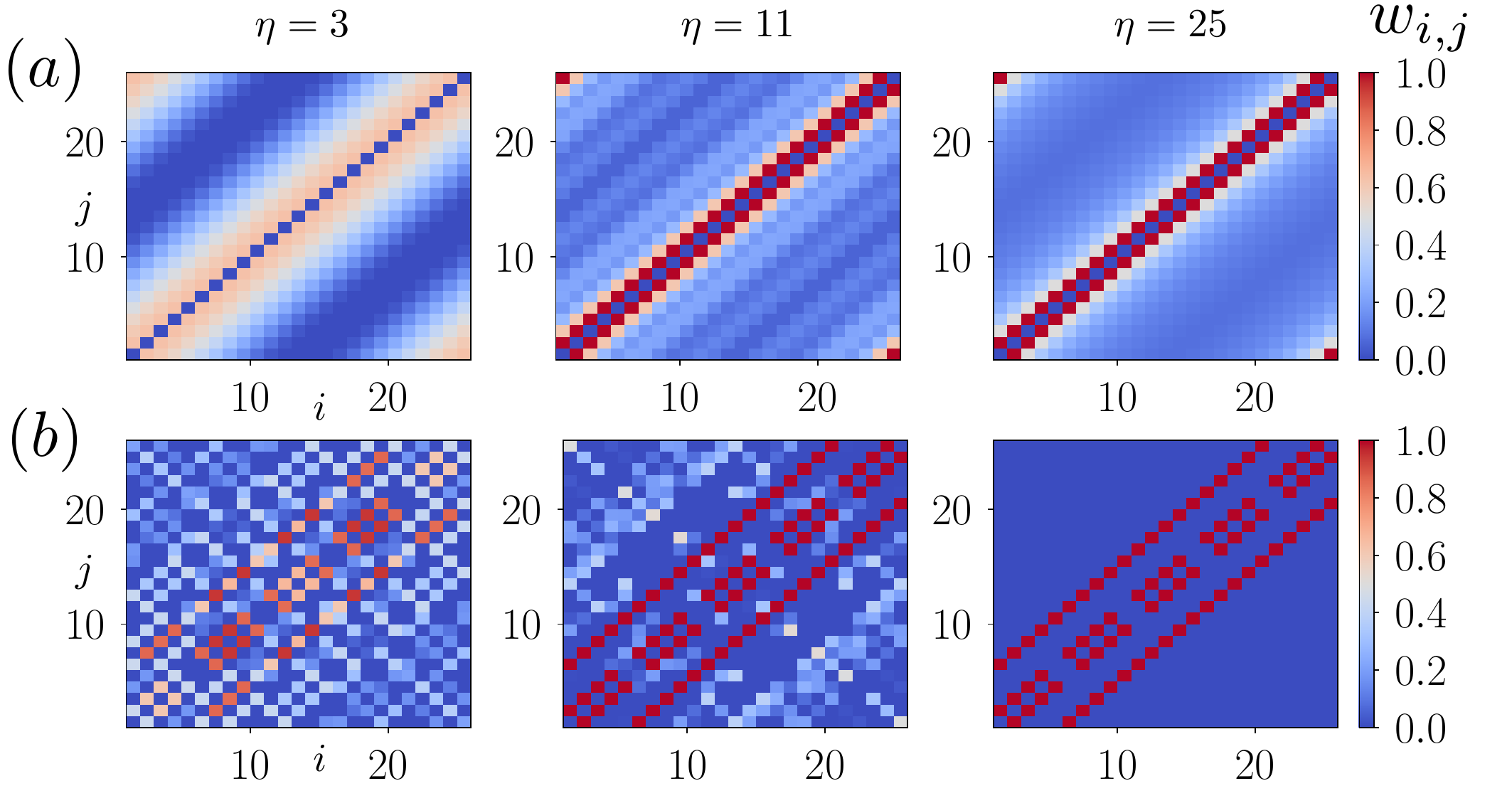}
\caption{\textit{Engineering of spin models.} 
(a) Long range interactions $w_{ij}=1/|i-j|$ and periodic boundary conditions.
(b) 2D nearest neighbor interactions with open boundary conditions. 
Here, the indices $i$ correspond to 2D indices $\mathbf{i}=(i_x,i_y)$ of a square of $5\times5$ sites using the convention $i=i_x+5 i_y$.
\label{fig:spinmodels}}
\end{figure}

\textit{Engineering of spin models.---}We now extend our discussion to the multi-qubit case $N>2$ and 
provide a recipe how to generate a targeted and scalable unitary
$W=\exp(-i \sum_{i<j} w_{ij} \sigma_i^z \sigma_j^z)$ 
with desired spin-spin interaction parameters 
$w_{ij}$. 
To this end, we consider a sequence $q=1,\dots,\eta$ of successive cycles
where for each stroboscopic cycle (labeled by $q$) we may apply different coupling amplitudes, i.e., $g_i\to g^{(q)}_{i}$. 
For example, this could be done by pulsing the amplitudes $A_{i,n}$ via microwave control \cite{harvey18, Royer2017}.
The evolution at the end of the sequence is then given by 
$U_{\eta}=\exp(-i t_{p} \sum_{i<j} J^{(\eta)}_{i,j}  \sigma_i^z \sigma_j^z)$, with 
$J^{(\eta)}_{i,j} = \sum_{q=1}^{\eta} g^{(q)}_{i}g^{(q)}_{j} / \omega_1,$
and $t_{\mathrm{run}}=\eta t_p$ being the total run time. 
A straightforward way to generate the desired unitary, i.e., to obtain 
$w_{ij}=J^{(\eta)}_{i,j}t_{p}$, consists in   
diagonalizing the target matrix as
$w_{ij}=\sum^N_{q=1} w_{q} u_{i,q} u_{j,q}$ in terms of real eigenvalues $w_{q}$ and real eigenstates $u_{i,q}$. 
This leads immediately to the condition $g^{(q)}_{i}=\sqrt{w_{q} \omega_1 / t_{p}} u_{i,q}$ to generate exactly $W$ within $\eta=N$ number of cycles, 
with $t_{p} \geq w_{q}/J_{\mathrm{max}}$, where $J_{\mathrm{max}}$ denotes the largest available spin-spin coupling \cite{eigenvalue}.
In other words, we can engineer efficiently arbitrary spin-spin interactions after a time $t_{\mathrm{run}}=N t_p$ which only scales \textit{linearly} with the number of qubits;
$t_{\mathrm{run}}=2N t_p$ in the presence of spin echo.
These aspects are illustrated in Fig.~\ref{fig:spinmodels}, where we provide examples for $N=25$ and both
(a) a 1D long-range spin model with power law decay $w_{ij}=1/|i-j|^\alpha$ ($\alpha=1$) and
(b) a 2D model with nearest neighbor interactions (NN). 
The latter demonstrates that our recipe allows for the realization of general 
spin models in any spatial dimension and geometry (using a simple one-dimensional physical setup).
For both models, we observe the progressive emergence of the target spin interaction with increasing values for $\eta$, reaching the exact matrix at $\eta=N$.
The case of a spin glass  with random interactions, and the convergence analysis with respect to $\eta/N$ are presented in \cite{SM}.

\begin{figure}
\includegraphics[width=\columnwidth]{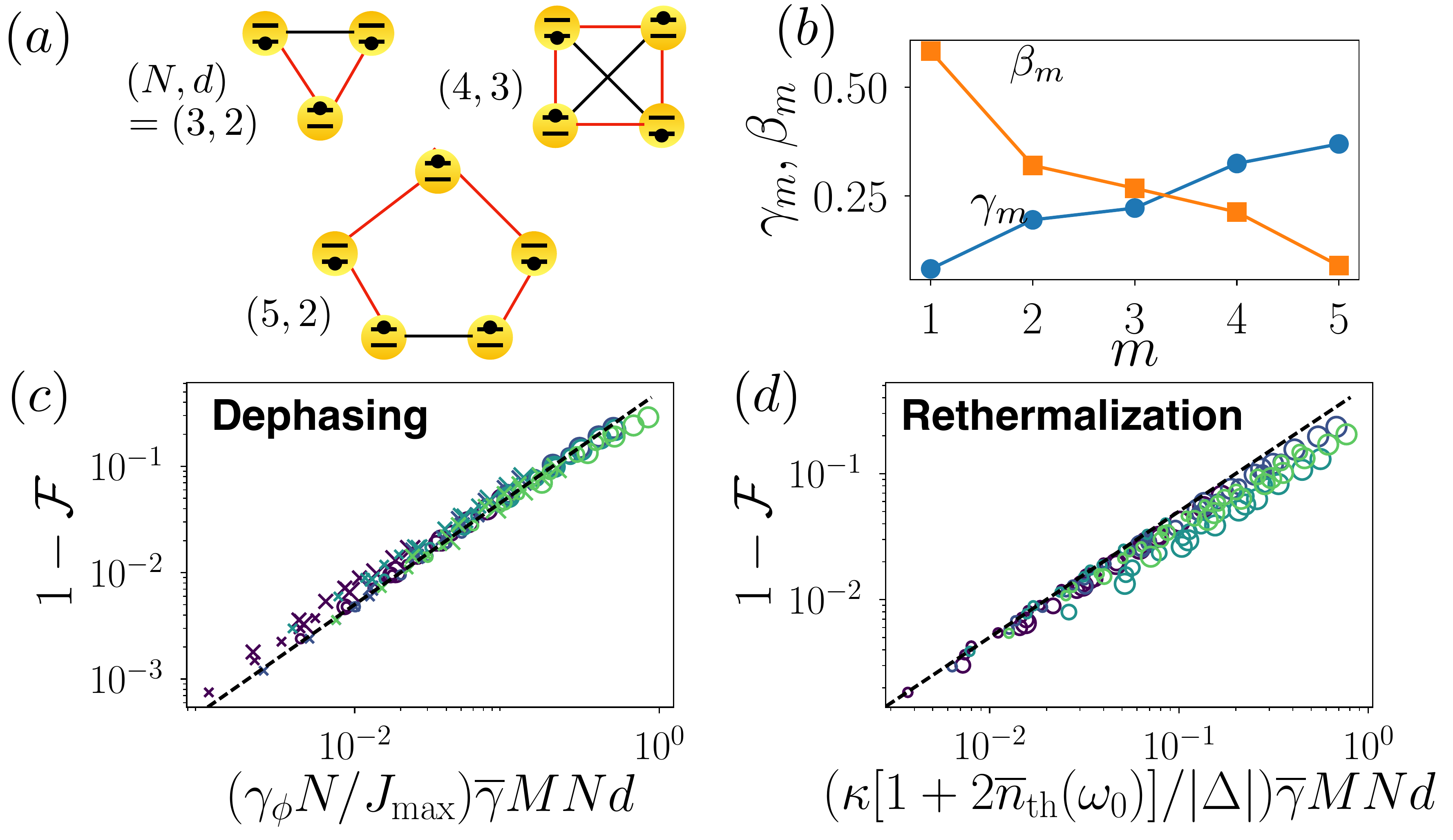}
\caption{{\it Simulation of QAOA for Max-Cut, in the presence of decoherence.} 
(a) $d$-regular graphs with $N=3,4,5$ used for our numerical analysis of decoherence. Our graph with $(N,d)=(6,4)$ is shown in Fig.~\ref{fig:setup}(c).
(b) Optimization parameters $\gamma_{m},\beta_{m}$ for $N=6$, $M=5$.
(c-d) Scaling of errors with respect to the optimized QAOA wave-function $\ket{\boldsymbol{\gamma},\boldsymbol{\beta}}$ for (c) dephasing and (d) rethermalization. 
For each panel, we consider the different graphs, depth $M=1,3,5$, $J_{\max}/|\Delta|=0.02,0.08$.
For (d), we consider $k_BT=0,\omega_{0}$. In (c-d), the dashed lines represents the curve $y=x/2$.
 \label{fig:QAOA}}
\end{figure}

\textit{QAOA.---}Finally, we show how to generalize the techniques outlined above in order to implement quantum algorithms that provide approximate solutions for hard combinatorial optimization problems such as Max-Cut [c.f. Fig.~\ref{fig:QAOA} and \cite{SM}].
As shown in Refs.\cite{farhi14, farhi16}, good approximate solutions to these kind of problems can be found by preparing the state 
$\left|\boldsymbol{\gamma}, \boldsymbol{\beta}\right> = U_{x}(\beta_{M})U_{zz}(\gamma_{M}) \cdots U_{x}(\beta_{1})U_{zz}(\gamma_{1})\left|s\right>$,  
with $U_x(\beta_m)=\exp[-i\beta_{m} \sum_{i} \sigma_{i}^{x}]$, and $U_{zz}(\gamma_m)=\exp[-i\gamma_m H_C]$, 
where $H_C$ is the cost Hamiltonian encoding the optimization problem, 
starting initially from a product of $\sigma^{x}$ eigenstates, i.e., $\left|s\right>=\left|-,-,\dots \right>$, with $\ket{-}=(\ket{0}-\ket{1})/\sqrt{2}$.  
In our scheme, this family of states can be prepared by alternating single-qubit operations $U_x(\beta_m$) with targeted spin-spin interactions generated as described above, with $W \to U_{zz}(\gamma_m)$.
Accordingly, for QAOA we repeat our spin-engineering recipe $M$-times with single-qubit rotations interspersed in between.
This preparation step is then followed by a measurement in the computational basis, giving a classical string $z$, 
with which one can evaluate the objective function $\langle H_C \rangle$
of the underlying combinatorial problem at hand. 
Repeating this procedure will provide an optimized string $z$, with 
the quality of the result improving as the depth of the quantum circuit $M$ is increased \cite{farhi14, farhi16}. 
To illustrate and verify this approach, we have numerically simulated QAOA with up to $N=6$ qubits solving Max-Cut for several $d$-regular graphs with weights $w_{i,j}=w_{i,j}^{(d)} +d \delta_{i,i}$, as depicted in Fig.~\ref{fig:QAOA}(a) and Fig.~\ref{fig:setup}(c), based on our model Hamiltonian given in Eq.(\ref{eq:model}), 
while accounting for both finite temperature and decoherence in the form of qubit dephasing and rethermalization of the resonator mode. 
While our general multi-mode setup should (in principle) be well suited for the implementation of QAOA, 
here (in order to allow for an exact numerical treatment) we consider a simplified single-mode problem (with resonator frequency $\omega_{0}$), 
as could be realized using the resonance condition introduced by a monochromatically modulated coupling \cite{harvey18, Royer2017}.
Specifically, we simulate the Hamiltonian 
$H=\sum_{i}(\omega_{i}/2) \sigma_{i}^{z} + \Delta a^{\dagger}a + \sum_{i}g_{i}\sigma_{i}^{z}\otimes(a+a^{\dagger})$
with controllable couplings $g_{i}$ \cite{harvey18, Royer2017}, detuning $\Delta = \omega_{0} - \Omega$ and $J_{ij}=-2g_{i}g_{j}/\Delta \leq J_{\mathrm{max}}$, 
supplemented by standard dissipators to account for 
(i) qubit dephasing on a timescale $\sim T_{2}=1/\gamma_\phi$ and 
(ii) rethermalization of the resonator mode with an effective decay rate $\sim \kappa \bar{n}_{\mathrm{th}}(\omega_{0})$ \cite{T1decay};
cf. \cite{SM} for further details. 
As demonstrated in Fig.~\ref{fig:setup}(c), for small-scale quantum systems (that are accessible to our exact numerical treatment) our protocol efficiently solves Max-Cut with a circuit depth of $M \lesssim 5$, finding the ground-state energy with very high accuracy (blue curve), corresponding to 4 \textit{cuts} (shown in red in the inset), 
even in the presence of moderate noise [compare the cross and plus symbols in Fig.~\ref{fig:setup}(c)].

\textit{Decoherence and implementation.---}Based on our numerical findings and further analytical arguments, we now turn to the eventual limitations imposed by decoherence. 
Here, we focus on the QAOA protocol, since both our 
(i) \textit{hot} gate (cf. \cite{SM} for a full decoherence-induced error analysis thereof) and 
(ii) the spin engineering protocol can be viewed as less demanding limits of QAOA, 
where either $M$ or $N$ (or both) are small, thereby yielding comparatively smaller errors because of a shorter run-time; for example, for the two-qubit phase gate $M=1$, $N=2$. 
The total QAOA run-time $t_{\mathrm{run}}$ can be upper-bounded as $t_{\mathrm{run}} \approx \overline{\gamma} M Nd/J_{\max}$, 
with $\overline{\gamma}=1/M \sum_{m} \gamma_{m}$ and the factor $Nd/J_{\max}$ corresponding to the (maximum) time required to implement all eigenvalues $w_q \lesssim d$ of the Max-Cut problem. 
To keep decoherence effects minimal, this timescale should be shorter than all relevant noise processes. 
The accumulated dephasing-induced error can be estimated as 
$\xi_{\phi} \sim \gamma_\phi N \times \overline{\gamma} M Nd/J_{\max}$, 
where $\sim \gamma_\phi N$ is the effective many-body dephasing rate (c.f. \cite{SM}); as shown in Fig.~\ref{fig:QAOA}(c), we have numerically confirmed this scaling for all graphs shown in panels Fig.~\ref{fig:QAOA}(a) and Fig.~\ref{fig:setup}(c). 
Similarly, as demonstrated in Fig.~\ref{fig:QAOA}(d), the indirect rethermalization-induced dephasing error, mediated by incoherent evolution of the resonator mode, can be quantified as 
$\xi_{\kappa}\sim \kappa_{\mathrm{eff}}\times \overline{\gamma} M Nd/|\Delta|$, with total linewidth $\kappa_{\mathrm{eff}} = \kappa(2\bar{n}_{\mathrm{th}}\left(\omega_{0}\right) + 1)$. 
The total decoherence-induced error $\xi = \xi_{\phi} + \xi_{\kappa}$ can then be optimized with respect to $\Delta$, yielding the compact expression 
$\xi \approx \overline{\gamma}dMN^{3/2}/\sqrt{C}$, with the cooperativity $C=g^2/(\gamma_\phi \kappa_{\mathrm{eff}})$. 
With this expression, we can bound the maximum number of qubits $N$ and circuit depth $M$ for a given physical setup with cooperativity $C$. 

Specifically, our scheme could be implemented
based on superconducting qubits or quantum-dot based qubits coupled by a common high-quality transmission line, with details given in \cite{SM}.
For concreteness, let us consider quantum-dot based qubits \cite{beaudoin16, jin12, schuetz17, harvey18, russ17}
where longitudinal coupling could be modulated via both the detuning \cite{harvey18} or inter-dot tunneling parameter \cite{jin12}, respectively. 
With projected two-qubit gate times of $\sim 10\mathrm{ns}$ \cite{harvey18, jin12}, a coherence time of $T_{2} \approx 10\mathrm{ms}$ \cite{veldhorst14, veldhorst15}, and 
$\omega_{0}/2\pi \approx  1\mathrm{GHz}$ with quality factor $Q\sim 10^6$ \cite{barends08, megrant12, bruno15}, 
we estimate decoherence errors to be small ($\lesssim 3\%$) for up to $N \approx 50$ qubits and a QAOA circuit depth of $M \approx 10$ for a graph with $d\approx4$, respectively, 
even in the presence of non-zero thermal occupation with $\bar{n}_{\mathrm{th}}\left(\omega_{0}\right) \approx 3$.
A similar analysis can be made for superconducting qubits \cite{SM}. 
Note that these estimates might be very conservative, as the essential figure of merit in QAOA is not the quantum state fidelity $\mathcal{F}$ but the probability to find the optimal (classical) bit-string $z$ in a sample of projective measurements $\{z_1,z_2,\dots\}$, which are obtained after many repetitions of the experiments.

\textit{Conclusion.---}To conclude, we have presented a protocol to generate fast, coherent, long-distance coupling between solid-state qubits, 
without any ground-state cooling requirements. 
While this approach has direct applications in terms of the engineering of spin models --- e.g. to implement quantum optimization algorithms --- it would be interesting to further develop our theoretical treatment in order to increase the level of robustness of our scheme, e.g. to apply protocols based on error correcting photonic codes~\cite{Michael2016}, which can protect against single photon losses or rethermalization.
Yet another interesting research direction would be to adapt our scheme to other physical setups, say solid-state defect centers coupled by phonons \cite{rabl10}.

\begin{acknowledgments}
\textit{Acknowledgments.---}We thank Shannon Harvey, Hannes Pichler, Pasquale Scarlino, Denis Vasilyev, Shengtao Wang and Leo Zhou for fruitful discussions.
Numerical simulations were performed using the ITensor library (http://itensor.org) and QuTiP~\cite{Johansson2013}. 
MJAS would like to thank the Humboldt foundation for financial support.
LMKV acknowledges support by an ERC Synergy grant (QC-Lab). 
JIC acknowledges the ERC Advanced Grant QENOCOBA under the EU Horizon 2020 program (grant agreement 742102).
Work in Innsbruck is supported by the ERC Synergy Grant UQUAM, the SFB FoQuS (FWF Project No. F4016-N23), and the Army Research
Laboratory Center for Distributed Quantum Information via the project SciNet.
Work at Harvard University was supported by NSF, Center for Ultracold Atoms, CIQM, Vannevar Bush Fellowship, AFOSR MURI and Max Planck Harvard Research Center for Quantum Optics.

M.J.A.S. and B.V. contributed equally to this work.

\end{acknowledgments}

\newpage \onecolumngrid \newpage { \center \bf \large  Supplemental Material for: \\  Quantum Simulation and Optimization in Hot Quantum Networks \vspace*{0.1cm}\\  \vspace*{0.0cm} } 
\begin{center} M.J.A. Schuetz,$^{1,*}$ B. Vermersch,$^{2,3,*}$ G. Kirchmair,$^{2,3}$ L.M.K. Vandersypen,$^{4}$ J.I. Cirac,$^{5}$ M.D. Lukin,$^{1}$  and P. Zoller$^{2,3}$ \\ 
\vspace*{0.15cm} 
\small{\textit{$^{1}$Physics Department, Harvard University, Cambridge, MA 02318,USA\\ 
$^{2}$Center for Quantum Physics, and Institute for Experimental Physics, University of Innsbruck, A-6020 Innsbruck, Austria\\ 
$^{3}$Institute for Quantum Optics and Quantum Information, Austrian Academy of Sciences, A-6020 Innsbruck, Austria\\ 
$^{4}$Institute for Experimental Physics, University of Innsbruck, A-6020 Innsbruck, Austria\\ 
$^{5}$QuTech and Kavli Institute of NanoScience, TU Delft, 
2600 GA Delft, The Netherlands}}\\
$^{6}$Max-Planck-Institut f\"ur Quantenoptik, Hans-Kopfermann-Str. 1, 85748 Garching, Germany\\ 
\vspace*{0.25cm} 
\end{center}

\twocolumngrid
\setcounter{section}{0}

\setcounter{equation}{0} 
\setcounter{figure}{0} 
\setcounter{table}{0} 
\setcounter{page}{1} 
\makeatletter 
\renewcommand{\theequation}{S\arabic{equation}} 
\renewcommand{\thefigure}{S\arabic{figure}} 
\renewcommand{\bibnumfmt}[1]{[S#1]} 
\renewcommand{\citenumfont}[1]{S#1} 


\section{Effective Spin-Spin Interactions} \label{Effective-spin-spin-interactions}
In this section, we analytically derive the expression for the effective coupling $J_{ij}=g_ig_j/\omega_1$, as presented in the main text (MT).

\textit{General results.---}We have introduced the effective spin-spin interaction as
\begin{equation}
J_{ij}=-2\sum^\infty_{n=1}\frac{g_{i,n}g_{j,n}}{\omega_{n}}, \label{eq:Jij_n}
\end{equation}
with the spin-resonator coupling parameters given as $g_{i,n}=g_i \sqrt{n} \int_{0}^L \cos (k_n x) f(x-x_i) dx$.
This yields
\begin{eqnarray}
J_{ij}&=&-\frac{2g_{i,n}g_{j,n}}{\omega_{1}} \int dx dx' f(x-x_i) f(x'-x_j) \times \nonumber \\
&& \times \sum^\infty_{n=1} \cos(k_n x) \cos(k_n x') \nonumber.
\end{eqnarray}
The sum over $n$ gives
\begin{eqnarray}
 &&\sum^\infty_{n=1} \cos(k_n x) \cos(k_n x') \nonumber \\
 &=&\frac{1}{2} \sum^\infty_{n=1} \cos(k_n(x-x'))+  \cos(k_n(x+x')) \nonumber \\
 &=&\frac{1}{4} \sum_{k\in \mathbb{Z}} \Big(\delta_{(x-x')/(2L)-k)}+\delta_{-(x+x')/(2L)-k)}\Big)-\frac{1}{2}\nonumber,
\end{eqnarray}
where we have used the Fourier Series decomposition of the Dirac comb 
\begin{eqnarray}
\sum_{n=1}^\infty\cos(k_n x)&=& \frac{1}{2}\left(\sum_{n=-\infty}^\infty e^{i n \pi x/L}-1\right)\nonumber \\
&=&\sum_{k\in \mathbb{Z}} \frac{1}{2}\left(   \delta(x/(2L)-k)-1\right).
\end{eqnarray}
Given the range of integration over $x,x'$, only the first Dirac $\delta$ function contributes to $J_{i,j}$. 
This leads to 
\begin{eqnarray}
J_{i,j}&=&\frac{g_i g_j}{\omega_1} \left( \int_0^L f(x-x_i) dx\int_0^L f(x-x_j) dx \right. \nonumber \\
&& \left.-L \int_0^L f(x-x_i) f(x-x_j)dx \right).
\end{eqnarray}
Note that in the standard situation $|x_{i} - x_{j}|\gg a$ ($a$ being the spatial extent of the function $f$), the second term is negligible.
Using the normalization property of $f(x-x_j)$, i.e., $\int_0^L f(x-x_i) dx = 1$, we arrive at the result presented in the main text. 


\textit{Box function.---}For a box function $f(x)=\frac{\delta_{x>0}\delta_{x<a}}{a}$, and assuming $|x_j-x_i|>a$ (and also the obvious condition $0<x_{i,j}<x_{i,j}+a<1$), the second term is exactly zero and we obtain 
$J_{i,j} = g_i g_j / \omega_1$, 
which does not depend on $a$, nor the qubit positions.

%

\section{Parametric Modulation of the Qubit-Resonator Coupling: Potential Advantages}

In this Appendix we discuss the possibility to potentially boost and fine-tune the effective spin-spin interactions $J_{ij}$
by parametrically modulating the longitudinal spin-resonator coupling. 

Specifically, consider the generalization of Eq.(\ref{eq:model}) with an off-resonant modulation of $g_{i,n}$ at the drive frequency $\Omega_{n}$,
i.e., $g_{i,n} \rightarrow g_{i,n}\left(t\right)=A_{i,n}\cos\left(\Omega_{n}t\right)$, with $\Delta_{n}=\omega_{n} - \Omega_{n}$
\footnote{If the driving amplitudes $A_{i,n}$ are zero for all but one specific mode, one recovers (approximately) a single-mode  problem \cite{harvey18SM, Royer2017SM}.}. 
When transforming to a suitable rotating frame and neglecting rapidly oscillating terms (in the limit $\Delta_{n}, A_{i,n} \ll \Omega_{n}$)
we obtain a time-independent Hamiltonian $H$ which maps directly onto the system studied
so far with the replacements $\omega_{n}\rightarrow \Delta_{n}$ and $g_{i,n}\rightarrow A_{i,n}/2$.
Accordingly, for stroboscopic times synchronized with the \textit{detuning} parameters $\Delta_{n}$ (where $\Delta_{n} \cdot t_{p}=2\pi p_{n}$ with $p_{n}$ integer) 
the unitary evolution in the \textit{lab} frame reduces to 
Eq.(\ref{eq:unitary-stroboscopic}), up to a free evolution term $\exp[-it_{p}\sum_{n}\Omega_{n}a_{n}^{\dagger}a_{n}]$ 
(which leaves the qubits untouched and even reduces to the identity as well if $\Omega_{n} \cdot t_{p}=2\pi q_{n}$ with $q_{n}$ integer),
with $J_{ij} \approx - \sum_{n} A_{i,n}A_{j,n} / 2\Delta_{n}$, the sign of which may be controlled by introducing relative phases between the driving terms \cite{harvey18SM, Royer2017SM}.

Provided that parametric modulation of the qubit-resonator coupling (discussed as extension (iii) in the main text) can be implemented, it comes with the following potential advantages: 
(1) Here, the commensurability condition applies to the (tunable) detuning parameters $\Delta_{n}=\omega_{n}-\Omega_{n}$ rather than the bare spectrum $\omega_{n}$. 
Therefore, even if the bare spectrum of the resonator $\omega_{n}$ is not commensurable, periodic disentanglement of the internal qubit degrees of freedom from the (hot) resonator modes can be achieved by choosing the driving frequencies $\Omega_{n}$ appropriately.
(2) The coupling $J_{ij}$ can be amplified by cranking up the classical amplitudes $A_{i,n}$, provided that $A_{i,n}\ll\Omega_{n}$ for self-consistency. 
Moreover, $J_{ij}$ is suppressed by the detuning $\Delta_{n}$ only (rather than the frequencies $\omega_{n}$ as is the case in the static scenario).
Still, the detuning should be sufficiently large in order to avoid photon-loss-induced dephasing \cite{Royer2017SM} and 
to keep the stroboscopic cycle time $\sim2\pi/\Delta_{n}$ sufficiently short; see below for quantitative, implementation-specific estimates. 
(3) Since the number of modes effectively contributing to $J_{ij}$ is well controlled by the choice $|A_{i,n}|\geq0$, the high-energy cut-off problem described above is very well-controlled.

\section{Timing-Induced Errors}

In this Appendix we analyze errors induced by timing inaccuracies.
Limited timing accuracy leads to deviations from the ideal stroboscopic
times $t_{p}$, with corresponding time jitter $\Delta t=t-t_{p}$.
For example, in quantum dot systems timing accuracies $\Delta t$
of a few picoseconds have been demonstrated experimentally \cite{bocquillon13}.
Here, we present analytical perturbative results that complement our
numerical results as presented and discussed in the main text. 

Our analysis starts out from the Hamiltonian given in Eq.(1) in the
main text. For notational convenience we rewrite this Hamiltonian
as 
\begin{equation}
H=\sum_{i}\frac{\omega_{i}}{2}\sigma_{i}^{z}+\sum_{n}\omega_{n}a_{n}^{\dagger}a_{n}+g\sum_{n}\mathcal{S}_{n}\otimes\left(a_{n}+a_{n}^{\dagger}\right),
\end{equation}
with $g\mathcal{S}_{n}=\sum_{i}g_{i,n}\sigma_{i}^{z}$. The time evolution
operator generated by this Hamiltonian reads in full generality 
\begin{eqnarray}
e^{-iHt} & = & U_{z}\left(\omega_{i}t/2\right)U_{zz}\left(J_{ij}t\right)W\left(t\right),\\
W\left(t\right) & = & U_{\mathrm{pol}}e^{-it\sum_{n}\omega_{n}a_{n}^{\dagger}a_{n}}U_{\mathrm{pol}}^{\dagger},
\end{eqnarray}
with the spin-dependent, multi-mode polaron transformation $U_{\mathrm{pol}}=\exp[\sum_{n}\left(g/\omega_{n}\right)\mathcal{S}_{n}\left(a_{n}-a_{n}^{\dagger}\right)]$,
as well as the single-qubit $U_{z}\left(\omega_{i}t/2\right)=\exp\left[-it\sum_{i}\left(\omega_{i}/2\right)\sigma_{i}^{z}\right]$,
and two-qubit gates $U_{zz}\left(J_{ij}t\right)=\exp[-it\sum_{i<j}J_{ij}\sigma_{i}^{z}\sigma_{j}^{z}]$,
respectively. While for stroboscopic times $W\left(t_{p}\right)=\mathds1$,
as discussed extensively in the main text, for non-stroboscopic times
($t=t_{p}+\Delta t$) generically $W\left(t\right)$ will entangle
the qubit and resonator degrees of freedom, with $W\left(t_{p}+\Delta t\right)=W\left(\Delta t\right)$,
thereby reducing the overall gate fidelity. 

Errors due to limited timing accuracy will come from two sources:
(i) First, as is the case for any unitary gate, there will be standard
errors in the realization of single and two-qubit gates coming from
limited timing control. For example, we can decompose the two-qubit
gate as $U_{zz}\left(J_{ij}t\right)=U_{zz}\left(J_{ij}\Delta t\right)U_{zz}\left(J_{ij}t_{p}\right)$,
where $U_{zz}\left(J_{ij}t_{p}\right)$ refers to the desired target
gate and $U_{zz}\left(J_{ij}\Delta t\right)$ results in undesired
contributions. The latter will be small provided that the random phase
angles are small, i.e., $\Delta\gamma_{ij}=J_{ij}\Delta t\ll1$. Accordingly,
the timing control $\Delta t$ has to be fast on the time-scale set
by the two-qubit interactions. A similar argument holds for the single
qubit gate $U_{z}\left(\omega_{i}t/2\right)$ which is assumed to
be controlled by spin-echo techniques. (ii) Second, for non-stroboscopic
times there will be errors due to the breakdown of the commensurability
condition (given by $\omega_{n}t_{p}=2\pi p_{n}$ with $p_{n}$ integer);
for non-stroboscopic times $W\left(t\right)$ does not simplify to
the identity matrix. This type of error is specific to our hot-gate
scheme. While all errors of type (i) are fully included in our numerical
calculations, within our analytical calculation presented here we
will focus on errors of type (ii), as these are specific to our (quantum-bus
based) hot gate approach. 

In the following we will focus on errors due to the breakdown of the
commensurability condition, as described by the unitary $W\left(\Delta t\right)$.
Using the relation $U_{\mathrm{pol}}a_{n}U_{\mathrm{pol}}^{\dagger}=a_{n}+\left(g/\omega_{n}\right)\mathcal{S}_{n}$,
we have 
\begin{equation}
W\left(\Delta t\right)=\exp\left[-i\Delta t\sum_{n}\omega_{n}(a_{n}^{\dagger}+\frac{g}{\omega_{n}}\mathcal{S}_{n})(a_{n}+\frac{g}{\omega_{n}}\mathcal{S}_{n})\right].
\end{equation}
The qubits are assumed to be initialized in a pure state, $\varrho\left(0\right)=\left|\psi\right\rangle _{0}\left\langle \psi\right|$.
In the absence of errors, ideally they evolve into the pure target
state defined as $\left|\psi_{\mathrm{tar}}\right\rangle =U_{z}\left(\omega_{i}t_{p}/2\right)U_{zz}\left(J_{ij}t_{p}\right)\left|\psi\right\rangle _{0}$,
which comprises both the single and two-qubit gates. As discussed
above, here we neglect standard errors of type (i) and set $\left|\psi_{\mathrm{tar}}\right\rangle =U_{z}\left(\omega_{i}t/2\right)U_{zz}\left(J_{ij}t\right)\left|\psi\right\rangle _{0}$
at time $t=t_{p}+\Delta t$, assuming that $\omega_{i}\Delta t,J_{ij}\Delta t\ll1$.
Initially, the resonator modes are assumed to be in a thermal state,
with $\rho_{\mathrm{th}}=\prod_{n}e^{-\beta\omega_{n}a_{n}^{\dagger}a_{n}}/Z_{n}$,
and $\beta=1/k_{B}T$. Then, the full evolution of the coupled spin-resonator
system reads 
\begin{eqnarray}
\rho\left(t\right) & = & e^{-iHt}\varrho\left(0\right)\otimes\rho_{\mathrm{th}}e^{iHt},\\
 & = & W\left(\Delta t\right)\left(\varrho\left(t\right)\otimes\rho_{\mathrm{th}}\right)W^{\dagger}\left(\Delta t\right),
\end{eqnarray}
where $\varrho\left(t\right)=\left|\psi_{\mathrm{tar}}\right\rangle \left\langle \psi_{\mathrm{tar}}\right|$
refers to the qubit's pure (target) density matrix at time $t$ in
the case of ideal, noise-free evolution, while $\rho\left(t\right)$
gives the density matrix of the coupled spin-resonator system in the
presence of errors caused by incommensurate timing. The fidelity of
our protocol is defined as 
\begin{equation}
\mathcal{F}\left(t\right)=\left<\psi_{\mathrm{tar}}|\mathrm{Tr}_{\mathrm{res}}\left[\rho\left(t\right)\right]|\psi_{\mathrm{tar}}\right>,
\end{equation}
where $\mathrm{Tr}_{\mathrm{res}}\left[\dots\right]$ denotes the
trace over the resonator degrees of freedom. In order to derive a
simple, analytical expression for the incommensurabiliy-induced error
$\xi_{\mathrm{timing}}=1-\mathcal{F}$, in the following we restrict
ourselves to a single mode, taken to be the mode $a_{1}$ (for small
errors similar error terms due to multiple incommensurate modes can
be added independently); also note that our complementary numerical
results cover the multi-mode problem. Next, we perform a Taylor expansion
of the undesired unitary as 
\begin{equation}
W\left(\Delta t\right)\approx\mathds1-i\mathcal{O}_{1}-\frac{1}{2}\mathcal{O}_{1}^{2},\label{eq:Taylor-expansion-error}
\end{equation}
with 
\begin{equation}
\mathcal{O}_{1}=\omega_{1}\Delta t\left[a_{1}^{\dagger}a_{1}+\frac{g}{\omega_{1}}\mathcal{S}_{1}\left(a_{1}+a_{1}^{\dagger}\right)+(\frac{g}{\omega_{1}})^{2}\mathcal{S}_{1}^{2}\right].
\end{equation}
This approximation is valid provided that the effective phase error
is sufficiently small, that is $\Delta\varphi=\omega_{1}\Delta t||a_{1}^{\dagger}a_{1}||\ll1$;
approximately $\Delta\varphi\approx\omega_{1}\Delta t\bar{n}_{\mathrm{th}}\left(\omega_{1}\right)$,
where $\bar{n}_{\mathrm{th}}\left(\omega_{1}\right)$ gives the thermal
occupation of the mismatched mode. Then, up to second order in $\Delta\varphi$,
we obtain 
\begin{eqnarray}
\rho\left(t\right) & \approx & \varrho\left(t\right)\otimes\rho_{\mathrm{th}}-i\left[\mathcal{O}_{1},\varrho\left(t\right)\otimes\rho_{\mathrm{th}}\right]\nonumber \\
 &  & +\mathscr{D}\left[\mathcal{O}_{1}\right]\varrho\left(t\right)\otimes\rho_{\mathrm{th}},
\end{eqnarray}
where $\mathscr{D}\left[\mathcal{O}_{1}\right]\rho=\mathcal{O}_{1}\rho\mathcal{O}_{1}^{\dagger}-\frac{1}{2}\left\{ \mathcal{O}_{1}^{\dagger}\mathcal{O}_{1},\rho\right\} $
denotes the standard dissipator of Lindblad form. When tracing out
the resonator degrees of freedom and computing the overlap with the
ideal qubit's target state $\left|\psi_{\mathrm{tar}}\right\rangle $,
the first order terms are readily shown to vanish, and the leading
order terms scale as $\sim\Delta t^{2}$ (in agreement with our numerical
results). Evaluating the second-order terms, we obtain a compact expression
for the error given by 
\begin{eqnarray}
\xi_{\mathrm{timing}} & \approx & \left(\omega_{1}\Delta t\right)^{2}\{\left(2\bar{n}_{\mathrm{th}}\left(\omega_{1}\right)+1\right)\left(g/\omega_{1}\right)^{2}\left(\Delta\mathcal{S}_{1}\right)^{2}\nonumber \\
 &  & +\left(g/\omega_{1}\right)^{4}\left(\Delta\mathcal{S}_{1}^{2}\right)^{2}\}.\label{eq:commensurability-error-wo-spin-echo}
\end{eqnarray}
Here, $\left(\Delta\mathcal{S}_{1}\right)^{2}=\left<\psi_{\mathrm{tar}}|\mathcal{S}_{1}^{2}|\psi_{\mathrm{tar}}\right>-\left<\psi_{\mathrm{tar}}|\mathcal{S}_{1}|\psi_{\mathrm{tar}}\right>^{2}$
denotes the variance of the collective spin-operator $\mathcal{S}_{1}$
in the ideal target state $\left|\psi_{\mathrm{tar}}\right\rangle $.
Typically, for $\bar{n}_{\mathrm{th}}\left(\omega_{1}\right)\gg1$
and $g/\omega_{1}\ll1$ the first term will dominate the overall error
and we obtain 
\begin{equation}
\xi_{\mathrm{timing}}\approx2\left(\omega_{1}\Delta t\right)^{2}\bar{n}_{\mathrm{th}}\left(\omega_{1}\right)\left(g/\omega_{1}\right)^{2}\left(\Delta\mathcal{S}_{1}\right)^{2}.
\end{equation}
While the error scales linearly with the thermal occupation $\bar{n}_{\mathrm{th}}\left(\omega_{1}\right)$,
it is suppressed quadratically for small phase errors $\omega_{1}\Delta t\ll1$
and weak spin-resonator coupling $g/\omega_{1}\ll1$. However, our
analytical calculation is valid only provided that the Taylor expansion
in Eq.(\ref{eq:Taylor-expansion-error}) is justified; again, this
is the case if $\Delta\varphi=\omega_{1}\Delta t||a_{1}^{\dagger}a_{1}||\ll1$
is satisfied. Still, our analytical treatment supports and complements
our numerical results in the three following ways: 
(i) The timing error is quadratic in the time jitter $\Delta t$, i.e., $\xi_{\mathrm{timing}}\sim\Delta t^{2}$.
(ii) The timing error is linearly proportional to the effective spin-spin
interaction $\sim J\sim g^{2}/\omega_{1}$; in agreement with our
numerical results, (in the absence of dephasing) timing errors are suppressed
for slow two-qubit gates. 
(iii) The timing error scales linearly with temperature $\sim k_{B}T/\omega_{1}$.

\section{Engineering of Spin Models}

In this Appendix we provide further details regarding the implementation of targeted, engineered spin models. 

Specifically, two more comments are in order: 
(i) For translation invariant models, the eigenstates of $w_{ij}$ can be written as sine and cosine waves with normalized
momentum $-\pi<k_q<\pi$. In particular for long-range models, we can obtain good approximations of $w_{ij}$ using only a restricted number $\eta<N$ of cycles corresponding to the lowest spatial frequencies $k_q$.
(ii) To satisfy the condition $w_q>0$, we can add to $w_{ij}$ a diagonal component $w_D\delta_{i,i}$, which does not contribute to the dynamics, and which can also be used to improve the convergence with $\eta$.

\section{Additional Numerical Results}

In this section, we present additional numerical results related to the realization of a phase gate between two distant qubits 
and the engineering of spin models (compare Figs.~2-3 of the main text). 

\begin{figure}
\includegraphics[width=\columnwidth]{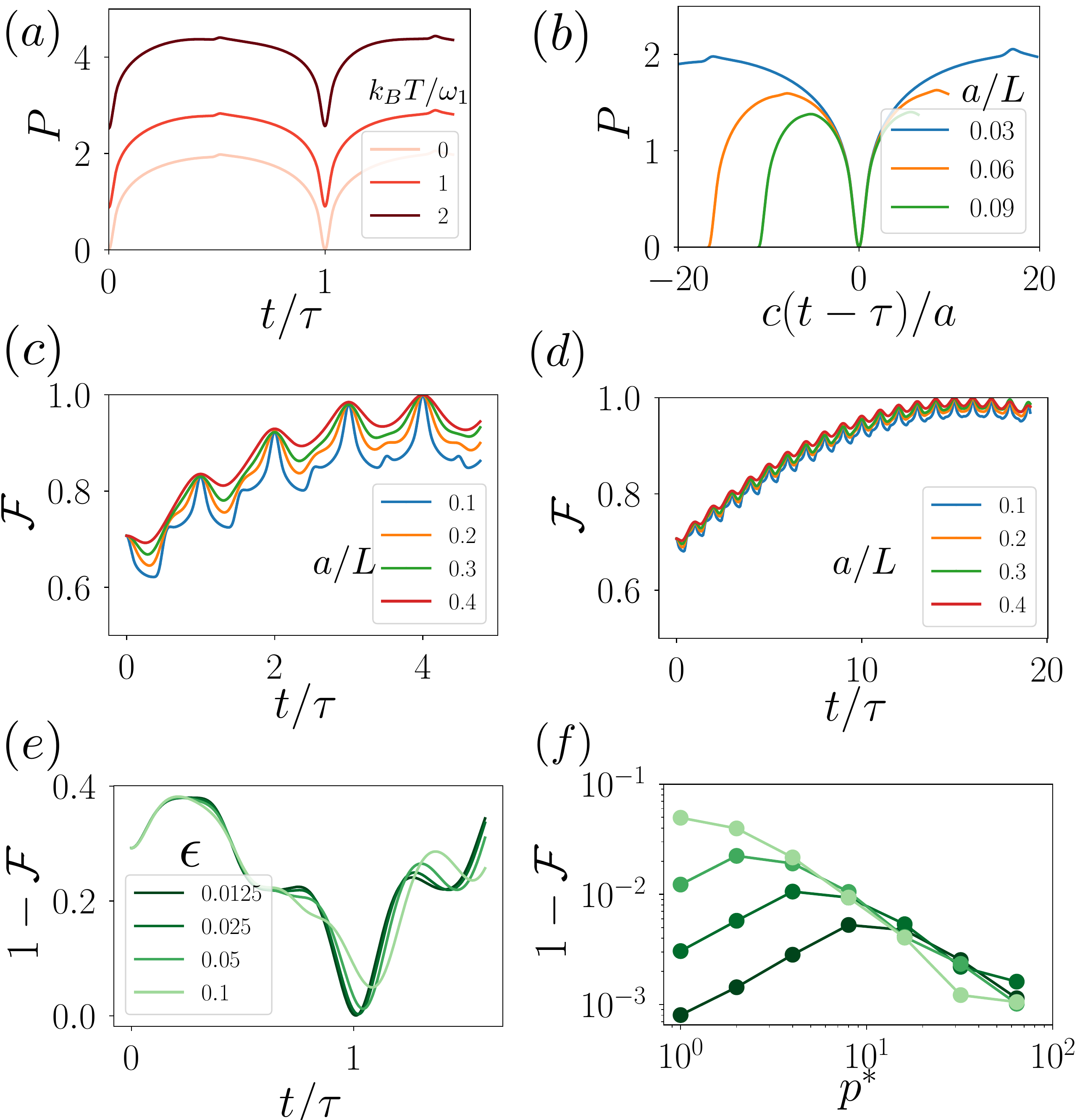}
\caption{{\it Hot phase gate between two distant qubits.} (a-b) Total photon number $P$ for the parameters of Fig.2(a) of the main text [panel (a)], and for $T=0$ and different cutoffs $a=0.03,0.06,0.09L$ [panel (b)].
(c-d) Fidelity $\mathcal{F}$ for smaller spin-resonator coupling parameters $g_{i}$, where the maximum fidelity is reached for $p^{\star}=4$ and $p^{\star}=16$ [panels (c) and (d), respectively]. These data  correspond to the analysis shown in Fig.2 (d) of the MT.  
(e)-(f) Gate error $1-\mathcal{F}$ in presence of a nonlinear term $\epsilon$ in the dispersion relation of the transmission line, for $p^\star=1$ versus time [panel (e)], and for different values of $p^\star$ at the optimal time when $1-\mathcal{F}$ is minimal [panel (f)]. Other parameters: $a=0.3L$, $T=0$.
\label{fig:photon-number}}
\end{figure}

\textit{Total photon number.---}The total photon number in the transmission line $P=\sum_n \langle a_n^\dag a_n\rangle $ is shown in Fig.~\ref{fig:photon-number}(a), 
for the parameters of Fig.~2(a) of the main text. 
At short times, the qubits excite a number of photons ($\sim 2$ for the chosen parameter set), which add up to the thermal background. 
These photons are then absorbed perfectly at the gate time $t=\tau$. 
As shown in panel (b), the number of emitted photons tends to slightly decrease with increasing values of $a$. 

\textit{Fidelity---}In panels (c-d) of Fig.~\ref{fig:photon-number} we provide further numerical results for spin-resonator coupling parameters $g_{i}$, 
where the maximum fidelity is reached for later times (rather than at $p^{\star}=1$, as discussed in the main text), namely for $p^{\star}=4$ and $p^{\star}=16$ [panels (c) and (d), respectively].
In all cases considered we take the ratio $g_{i}/\omega_{1}$ such that a maximally entangling gate can (in principle) be achieved at $p=p^{\star}$. 
Taking $g_{1}=g_{2}=g$, this is the case for $J_{12} t_{p^{\star}} = 2\pi (g/\omega_{1})^2 p^{\star} = \pi / 4$, as required for a maximally entangling gate of the form 
$U_{\mathrm{max}} = \exp[-i(\pi/4)\sigma_{1}^{z} \sigma_{2}^{z}]$. 
Since $p^{\star}$ can only take on integer values, the value of $g_{i}/\omega_{1}$ needs to be fine-tuned in order to achieve a maximally entangling gate; 
without fine-tuning generically the target state will still be entangled (but not maximally entangled, even in the absence of noise). 
As shown in panels (c-d) of Fig.~\ref{fig:photon-number}, periodic stroboscopic cycles for integer values of $p$ can clearly be identified. 
For values $p^{\star}\gg1$, many, small amplitude oscillations occur before the fidelity reaches its maximum value at the nominal gate time $t_{g}=\pi/(4J_{12})$. 
In this parameter regime, the effective dynamics for $\mathcal{F}(t)$ typically feature a slow (secular), large amplitude with high-frequency, small amplitude oscillations on top; 
therefore, the relevant timescale for timing errors (due to timing inaccuracies $\Delta t =t - t_{g}$) is set by the interaction as $\sim 1/J_{12}$, 
as exemplified in Fig.~\ref{fig:photon-number} (d) for $p^{\star}=16\gg1$.
Since the essential dynamics appear on a long timescale $\sim 1/J_{12}$, with only small changes occurring in the vicinity of $p^{\star}$, the constraints on timing errors are strongly relaxed, because stroboscopic precision on a timescale $\sim 1/\omega_{n}$ is not required in order to achieve a high-fidelity gate. 
Conversely, high-fidelity results can already be found in the parameter regime $\Delta t \ll \pi/4J_{12}$. 

\textit{Nonlinear spectrum.---}Next, we study potential errors due to a non-linear photonic spectrum (where $\omega_{n} \neq n \omega_{1}$). 
Before presenting our detailed numerical results, some general comments are in order: 
(i) First, note that this type of error can only occur in the multi-mode setup, but is entirely absent in the single-mode regime, 
as could be (approximately) realized using parametric modulation of the qubit-resonator coupling \cite{harvey18SM, Royer2017SM}. 
(ii) Second, the commensurability condition, as specified in the main text for a linear spectrum, can be generalized to spectra for which one can find 
a stroboscopic time $t^{\star}>0$ (and integer multiples thereof), for which $\omega_{1} t^{\star} = 2\pi p_{1}$, $\omega_{2} t^{\star} = 2\pi p_{2}$ etc.
can be satisfied for integer values $p_{1}, p_{2}, \dots$. 
This means that all fractions $\omega_{m} / \omega_{n} = p_{m} / p_{n}$ need to be rational numbers. 
Taking the ordering $\omega_{1} \leq \omega_{2} \leq \dots$, we may summarize these conditions as $\omega_{n}/\omega_{1} = p_{n} / p_{1} \in \mathbb{Q}$.
Then, with $\omega_{1} t^{\star} = 2\pi p_{1}$ satisfied, all remaining equations can be deduced as 
$\omega_{n} t^{\star} = (p_{n}/p_{1}) \omega_{1} t^{\star} = (p_{n}/p_{1}) 2\pi p_{1} = 2\pi p_{n}$. 
Therefore, given a specific spectrum $\omega_{n}$, (in principle) one may still find specific (stroboscopic) times $t^{\star}>0$ (and integer multiples thereof), 
for which the qubits disentangle entirely from the resonator modes, even if the spectrum is non-linear. 

Our numerical results can be found in Fig.~\ref{fig:photon-number}(e-f); here, we study the role of a nonlinear term in the dispersion relation of the transmission line, 
$\omega_n\to \omega_n=\omega_1 n - \delta \omega _n$, where (for concreteness) we consider a quadratic term of the form $\delta \omega_n=\epsilon \omega_1 (n-1)^2$.
In panel (e), we represent the gate error versus time for $p^{\star}=1$ and different values of $\epsilon$ (see legend). 
Around the gate time, the modes only partially synchronize, implying a minimal gate error which increases with $\epsilon$. 
We further quantify these effects by representing in panel (f) the gate error (at such optimal time) as a function of $p^{\star}$, for the same values of $\epsilon$. 
One clearly distinguishes two limits corresponding to $\delta \omega_n t_p \ll 1$ (resp. $\gg 1$), which we can both understand analytically, considering for simplicity the effect of the asynchronicity of the mode $n=2$ ($n=1$ is not affected by $\epsilon$), and $T=0$.
First, in the perturbative limit  $\delta \omega_2 t_p \ll 1$, the effect of the nonlinear term is analog to a timing error as discussed above, with the mode asynchronicity $\delta \omega_2 t_{p^{\star}}$ replacing the timing error $\omega_1 \Delta t$ in the expression of $W(t)$.  
This corresponds to a gate error
\begin{equation}
\xi_\mathrm{nonlinear}\approx  (\delta \omega_2 t_{p^{\star}})^2 (g/\omega_2)^2  (\Delta \mathcal{S}_2)^2, 
\end{equation}
scaling thus as $\epsilon^2 p^{\star}$, as confirmed by our numerical simulations. 
In the opposite limit, $\delta \omega_2 t_{p^{\star}} \gg 1$, the mode asynchronicity hits a maximum value $\delta \omega_2 t_{p^{\star}} (2\pi)\sim  \pi$, and the error reads
\begin{equation}
\xi_\mathrm{nonlinear}\approx \pi^2 (g/\omega_2)^2 (\Delta \mathcal{S}_2)^2, 
\end{equation}
scaling as $1/p^{\star}$, independently of $\epsilon$, as also seen in our numerical simulations. 
This means that, along the lines of timing errors, the effect of nonlinear terms can be reduced by increasing $p^{\star}$.

\begin{figure}
\includegraphics[width=\columnwidth]{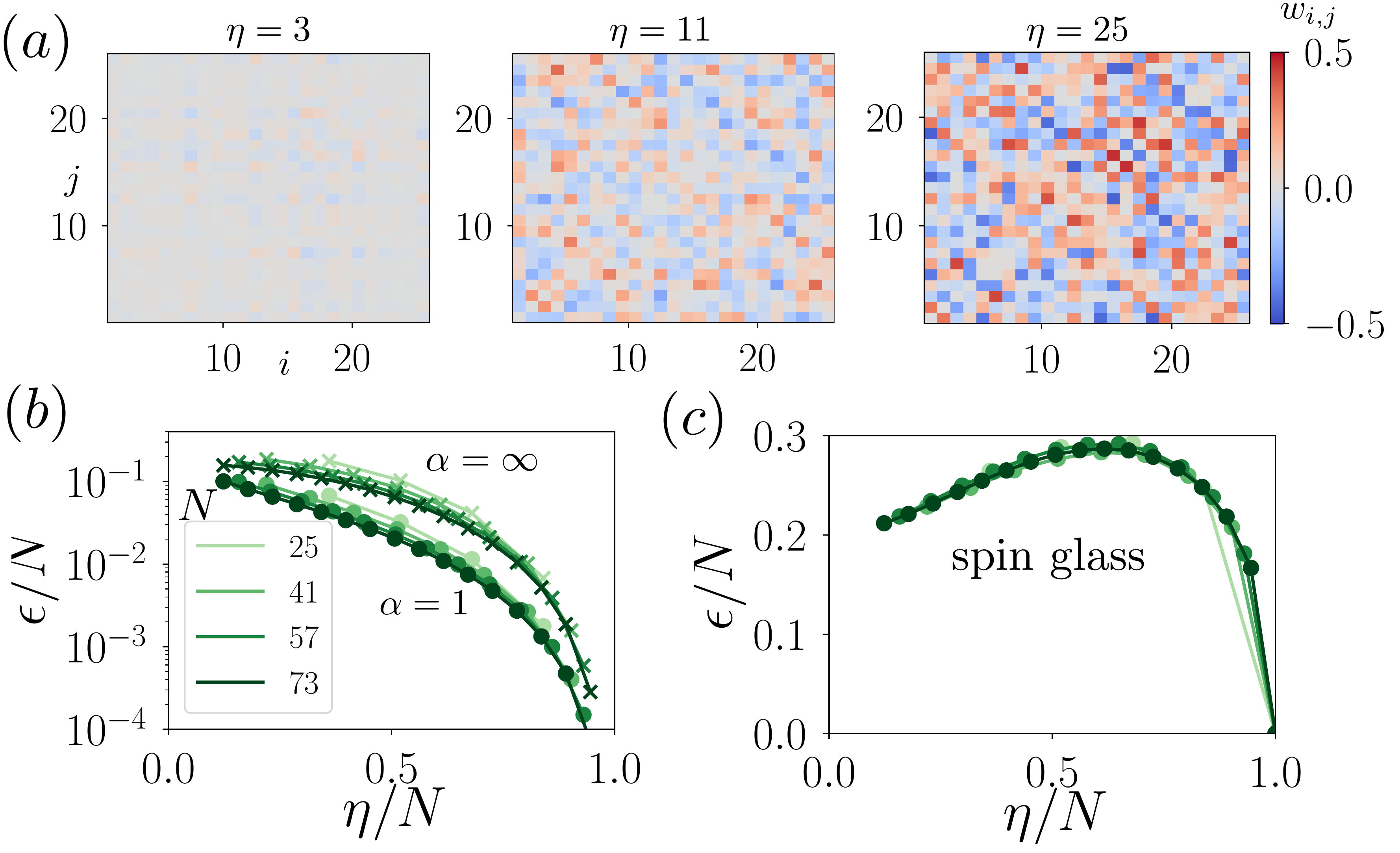}
\caption{{\it Engineering of spin models.} 
(a) Same as Fig.~3 MT for a spin glass with random interactions $w_{ij}$ between $[-0.5,0.5]$.
(b-c) Convergence analysis where we plot the error 
$\epsilon\equiv ||w_{ij}^{(\eta)}-w_{ij}||_2$ 
versus $\eta$ and different values of $N$.
\label{fig:SpinModels}}
\end{figure}

\textit{Engineering of spin models.---}In Fig.~\ref{fig:SpinModels}, we present additional numerical results on the engineering of spin models. 
In panel (a), we represent the formation of a spin glass with random interactions. 
In contrast to the models presented in Fig.~3 MT, one requires to implement the full spectrum, i.e., to use $\eta=N$, to obtain a faithful generation of the target matrix.
The convergence of the generated matrix $w_{ij}$ with $\eta/N$ is shown in Fig.~\ref{fig:SpinModels}(b) for 1D models with nearest neighbor interactions 
and with power law decay $\alpha=1$.
In both cases, we obtain a good representation of the targeted interactions for $\eta \gtrsim N/2$.
Note that the convergence to NN interactions occurs at later times compared to the power-law case  due to high spatial frequencies in the spectrum. 
As already shown in panel (a), to obtain a true spin glass model, one instead requires to implement the full spectrum of $W$, see Fig.~\ref{fig:SpinModels}(c).

\section{Decoherence Analysis}

In this section, we provide detailed background material related to effects due to decoherence. 
First, we present the Master equation used in order to model decoherence in the form of qubit dephasing and resonator rethermalization. 
Next, we analytically derive an expression for the gate error caused by qubit dephasing. 
Thereafter, we numerically analyze rethermalization-induced errors. 
Finally, we show that the total error due to both (i) dephasing and (ii) rethermalization can be quantified in terms of a single cooperativity parameter. 

\subsection{Master Equation}

\textit{Master equation.}---Within a standard Born-Markov approach,
the noise processes described above can be accounted for by a master
equation for the system's density matrix $\rho$ as 
\begin{eqnarray}
\dot{\rho} & = & -i\left[H_{\mathrm{id}},\rho\right]+(\gamma_{\phi}/2)\sum_{i}\mathscr{D}\left[\sigma_{i}^{z}\right]\rho\\
 &  & +\sum_{n}\kappa_{n}\left(\bar{n}_{\mathrm{th}}\left(\omega_{n}\right)+1\right)\mathscr{D}\left[a_{n}\right]\rho\\
 &  & +\sum_{n}\kappa_{n}\bar{n}_{\mathrm{th}}\left(\omega_{n}\right)\mathscr{D}\left[a_{n}^{\dagger}\right]\rho,
\end{eqnarray}
where $H_{\mathrm{id}}$ describes the ideal (error-free), coherent
evolution for longitudinal coupling between the qubits and the resonator
mode, and $\gamma_{\phi}=1/T_{2}$ is the pure dephasing rate. The second
and third line describe rethermalization of the modes $a_{n}$ towards
the a thermal state with an effective rate $\sim\kappa_{n}\left(\bar{n}_{\mathrm{th}}\left(\omega_{n}\right)+1\right)\approx k_{B}T/Q_{n}$. 

This simple noise model is valid within the so-called approximation
of independent rates of variation \cite{cohen-tannoudji92}, where
the interactions with the environment are treated separately for spin
and resonator degrees of freedom; in other words, they can approximately
treated as independent entities and the terms (rates of variation)
due to internal and dissipative dynamics are added independently.
While for ultra-strong coupling the qubit-resonator system needs to
be treated as a whole when studying its interaction with the environment
\cite{beaudoin11}, yielding irreversible dynamics through jumps between
dressed states (rather than bare states), in the weak coupling regime
$\left(g_{i,n}\ll\omega_{n}\right)$ one can resort to the standard
(quantum optical) dissipators given above, with $\mathscr{D}\left[a\right]\rho=a\rho a^{\dagger}-1/2\left\{ a^{\dagger}a,\rho\right\} $. 

\subsection{Dephasing-Induced Errors}

\textit{Dephasing-Induced Errors.}---In this Appendix we provide an analytical
model for dephasing-induced errors. Neglecting rethermalization-induced
errors for the moment, here we consider the following Master equation
\begin{equation}
\dot{\rho}=\underset{\mathcal{L}_{0}\rho}{\underbrace{-i\left[H_{\mathrm{id}},\rho\right]}}+\underset{\mathcal{L}_{1}\rho}{\underbrace{(\gamma_{\phi}/2)\sum_{i}\mathscr{D}\left[\sigma_{i}^{z}\right]\rho}},\label{eq:Master-equation-pure-dephasing-analytical-1}
\end{equation}
where $H_{\mathrm{id}}$ describes the ideal (error-free), coherent
evolution for longitudinal coupling between the qubits and the resonator
mode, and $\gamma_{\phi}$ is the pure dephasing rate. Since the super-operators
$\mathcal{L}_{0}$ and $\mathcal{L}_{1}$ as defined in Eq.(\ref{eq:Master-equation-pure-dephasing-analytical-1})
commute, that is $\left[\mathcal{L}_{0},\mathcal{L}_{1}\right]=0$
(since $\left[H_{\mathrm{id}},\mathscr{D}\left[\sigma_{i}^{z}\right]X\right]=\mathscr{D}\left[\sigma_{i}^{z}\right]\left[H_{\mathrm{id}},X\right]$
for any operator $X$), the full evolution simplifies to 
\begin{equation}
\rho\left(t\right)=e^{\mathcal{L}_{1}t}e^{\mathcal{L}_{0}t}\rho\left(0\right)=e^{\mathcal{L}_{1}t}\rho_{\mathrm{id}}\left(t\right),
\end{equation}
where we have defined the ideal target state at time $t$ as $\rho_{\mathrm{id}}\left(t\right)=\exp\left[\mathcal{L}_{0}t\right]\rho\left(0\right)$,
which, starting from the initial state $\rho\left(0\right)$, exclusively
accounts for the ideal (error-free), coherent evolution. For small
infidelities $\left(\gamma_{\phi}t\ll1\right)$, the deviation from
the ideal dynamics $\Delta\rho=\rho-\rho_{\mathrm{id}}$ is approximately
given by 
\begin{equation}
\Delta\rho\left(t\right)\approx\gamma_{\phi}t/2 \sum_{i}\mathscr{D}\left[\sigma_{i}^{z}\right]\rho_{\mathrm{id}}\left(t\right),\label{eq:linear-dephasing-error-analytical}
\end{equation}
showing that (in the regime of interest where $\gamma_{\phi}t\ll1$)
the dominant dephasing induced errors are linearly proportional to
$\sim\gamma_{\phi}t_{g}\sim\gamma_{\phi}/J_{ij}$, as expected; here,
$t_{g}\sim1/J_{ij}$ is the relevant gate time which has to be short
compared to $\gamma_{\phi}^{-1}$. 

In the following we compute the dephasing-induced error analytically.
We define the pure qubit target state as $\left|\Psi_{\mathrm{tar}}\right\rangle $
and take the state fidelity $\mathcal{F}$ as our figure of merit,
with 
\begin{equation}
\mathcal{F}\left(t_{g}\right)=\left<\Psi_{\mathrm{tar}}|\rho_{q}\left(t_{g}\right)|\Psi_{\mathrm{tar}}\right>,
\end{equation}
where $\rho_{q}\left(t_{g}\right)=\mathrm{Tr}_{\mathrm{res}}\left[\rho\left(t_{g}\right)\right]=\mathrm{Tr}_{\mathrm{res}}\left[\exp\left[\mathcal{L}t_{g}\right]\rho\left(0\right)\right]$
is the state of the qubits at time $t_{g}$, with $\mathcal{L}=\mathcal{L}_{0}+\mathcal{L}_{1}$
and $\mathrm{Tr}_{\mathrm{res}}\left[\dots\right]$ denoting the trace
over the resonator degrees of freedom. Since the qubits ideally disentangle
from the resonator modes for stroboscopic times and since $\mathcal{L}_{1}$
acts on the qubit degrees of freedom only, we find 
\begin{eqnarray}
\rho_{q}\left(t_{g}\right) & = & e^{\mathcal{L}_{1}t_{g}}\mathrm{Tr}_{\mathrm{res}}\left[e^{\mathcal{L}_{0}t_{g}}\left|\Psi_{0}\right\rangle \left\langle \Psi_{0}\right|\otimes\rho_{\mathrm{th}}\left(0\right)\right],\\
 & = & e^{\mathcal{L}_{1}t_{g}}\mathrm{Tr}_{\mathrm{res}}\left[\left|\Psi_{\mathrm{tar}}\right\rangle \left\langle \Psi_{\mathrm{tar}}\right|\otimes\rho_{\mathrm{th}}\left(0\right)\right],\\
 & = & e^{\mathcal{L}_{1}t_{g}}\left|\Psi_{\mathrm{tar}}\right\rangle \left\langle \Psi_{\mathrm{tar}}\right|.
\end{eqnarray}
The fidelity $\mathcal{F}\left(t_{g}\right)$ can then be expressed
\begin{equation}
\mathcal{F}\left(t_{g}\right)=\left<\Psi_{\mathrm{tar}}|e^{\mathcal{L}_{1}t_{g}}(\left|\Psi_{\mathrm{tar}}\right\rangle \left\langle \Psi_{\mathrm{tar}}\right|)|\Psi_{\mathrm{tar}}\right>.
\end{equation}
In the regime of interest (with small infidelities) we can approximate
the error $\xi\left(t_{g}\right)=1-\mathcal{F}\left(t_{g}\right)$
as 
\begin{equation}
\xi\left(t_{g}\right)=-t_{g}\left<\Psi_{\mathrm{tar}}|\mathcal{L}_{1}(\left|\Psi_{\mathrm{tar}}\right\rangle \left\langle \Psi_{\mathrm{tar}}\right|)|\Psi_{\mathrm{tar}}\right>
\end{equation}
With $\mathcal{L}_{1}$ as defined in Eq.(\ref{eq:Master-equation-pure-dephasing-analytical-1})
this leads to the compact expression 
\begin{equation}
\xi\left(t_{g}\right)=\gamma_{\phi}t_{g}/2\sum_{i}\left\{ 1-\left|\left<\Psi_{\mathrm{tar}}|\sigma_{i}^{z}|\Psi_{\mathrm{tar}}\right>\right|^{2}\right\} .
\end{equation}
Accordingly we only need to evaluate the expectation values of $\sigma_{i}^{z}$
in the ideal target state in order to estimate the dephasing-induced
fidelity error. Specifically for 
$\left|\Psi_{\mathrm{tar}}\right\rangle =\exp[-i t_{g}\sum_{i<j}J_{ij}\sigma_{i}^{z}\sigma_{j}^{z}]\left|\Psi_{0}\right\rangle $
it is sufficient to compute the expectation values of $\sigma_{i}^{z}$
in the initial state $\left|\Psi_{0}\right\rangle $, because 
\begin{equation}
\xi\left(t_{g}\right)=\gamma_{\phi}t_{g}/2\sum_{i}\left\{ 1-\left|\left<\Psi_{0}|\sigma_{i}^{z}|\Psi_{0}\right>\right|^{2}\right\} .
\end{equation}
For qubits initialized in the $x-y$ plane, 
e.g., $\left|\Psi_{0}\right\rangle = \otimes_{j=1}^{N}(\left|0\right\rangle _{j}+i\left|1\right\rangle _{j})/\sqrt{2}$,
the expectation values $\left<\Psi_{0}|\sigma_{i}^{z}|\Psi_{0}\right>=0$
vanish and we arrive at a (conservative) estimate of
\begin{equation}
\xi\left(t_{g}\right) \approx (N\gamma_{\phi}/2) t_{g}, \label{eq:error_dephasing}
\end{equation}
with $N=\sum_{i}$ being the number of qubits, and $\gamma_{\mathrm{eff}} = N\gamma_{\phi}/2$ describing the effective many-body dephasing rate. 
As expected the error
grows linearly with the gate time $\sim t_{g}/T_{2}$. 

\begin{figure}
\includegraphics[width=\columnwidth]{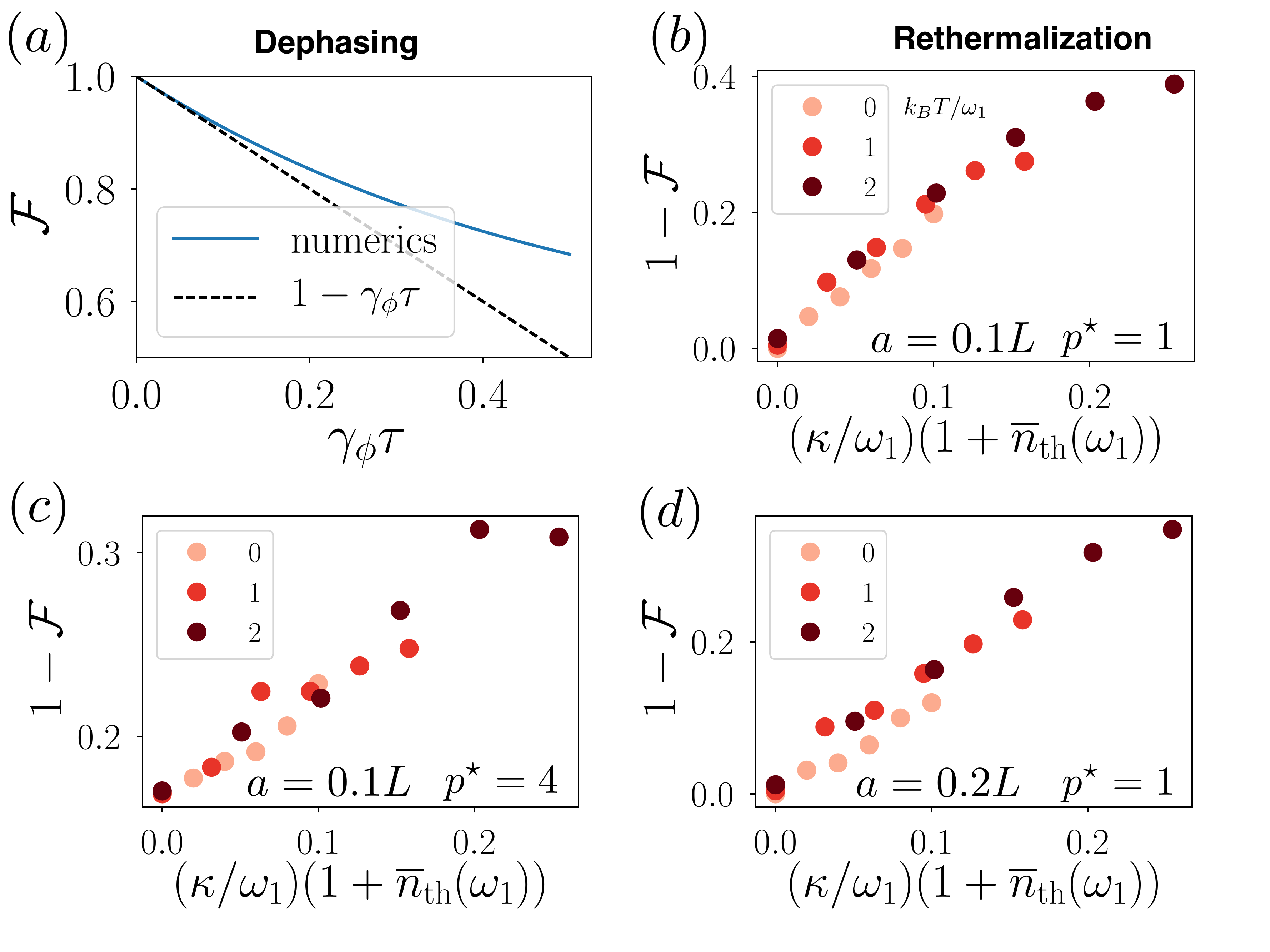}
\caption{{\it Dephasing and rethermalization induced errors.} 
(a) Effect of dephasing for the phase gate presented in Fig.~2 MT, as a function of the dephasing rate $\gamma_\phi$.
(b-d) Rethermalization induced error $\xi=1-\mathcal{F}$ due to coupling of the ($\sim 30$) resonator modes to a thermal reservoir, 
for three different temperatures (light red to dark red; see legend), two different cutoff values ($a/L=0.1, 0.2$; see text in panels), and different qubit-photon coupling parameters $g_{i}=g$. 
The latter are set as $g/\omega_{1}=1/\sqrt{8}$ (panels b,d) and $g/\omega_{1}=1/\sqrt{32}$ (panels c), respectively. 
In the small error regime of interest, the (linear) temperature dependence is well captured by the thermal occupation factor $\bar{n}_{\mathrm{th}}(\omega_{1})$, while the error is found to be independent of the coupling $g$.     
To simplify the numerical treatment, we considered a value $\kappa_n=\kappa$ independent of $n$.
\label{fig:Deco}}
\end{figure}

{\it Numerical verifications.---}First, as demonstrated in Fig.~\ref{fig:Deco}(a), we have numerically verified the liner error scaling [compare Eq.~\eqref{eq:error_dephasing}] 
induced by dephasing for $N=2$ qubits and a gate time $t_g=\tau$.
Second, we have numerically verified the scaling of the state error
\begin{equation}
\xi(t_\mathrm{run})\approx (\gamma_\phi N/2J_{\max})\bar{\gamma}MNd,
\end{equation}
for the modulated scheme applied to QAOA, see Fig. 4 MT. This is a direct consequence of Eq.~\eqref{eq:error_dephasing}, obtained for a total run time $t_g\to t_\mathrm{run}\sim \bar{\gamma}MNd/J_{\max}$ (see text). 

\subsection{Rethermalization-Induced Errors}

\textit{Errors for a two-qubit gate.---}We have first numerically verified that rethermalization-induced errors are independent of the qubit-resonator coupling strength $g_{i}$, as demonstrated in Fig.~\ref{fig:Deco}(b-d). In this case, we took into account the effect of decoherence by calculating the evolution of $100$ MPS quantum trajectories~\cite{Daley2014}. This finding can be understood from the fact that photon rethermalization leads to qubit dephasing (due to leakage of which-way information) at an effective rate $\sim g^2$ that scales quadratically with the qubit-dependent separation in phase space (i.e., the displacement amplitude), while the relevant gate time scales as $\sim 1/g^2$ \cite{Royer2017SM, schuetz17SM, rabl10SM}.  
When multiplying these two factors to obtain the effective error the dependence on $g$ drops out, leading to an effective error that is independent of $g$, as numerically verified in Fig.~\ref{fig:Deco}(b-d). Finally, for the two values of $a$ considered, we did not observe a significant effect of the cut-off value $a$ on rethermalization errors. 

{\it Scaling analysis for QAOA.---}We now consider the multi-qubit scenario. 
In Fig.~4(b) of the MT, we show a scaling analysis for QAOA in the single-mode case, which indicates that the total error can be estimated by
\begin{equation}
\xi(t_\mathrm{run}) \approx (\kappa (1+2\bar{n}_\mathrm{th}(\omega_0))/|\Delta|) \bar{\gamma}MNd.\label{eq:error_rethermalization}
\end{equation}
In order to interpret this numerical result, we first estimate the error accumulated during a cycle of duration $t_p$ implementing the component $q$ of the Hamiltonian $H_C$ (see MT). Following Refs.~\cite{Royer2017SM, schuetz17SM}, this corresponds to an error 
\begin{equation}
\xi(t_p) \approx (\kappa (1+2\bar{n}_\mathrm{th}(\omega_0))\bra{\psi}\mathscr{D}[S]\ket{\psi} t_p, 
\end{equation}
with $\ket{\psi}$ denoting the (ideal) target state obtained in the absence of noise ($\kappa=0$), 
the collective spin operator $S=\sum_i (g_i/\Delta) \sigma_i^z$, 
the spin-resonator coupling $g_i=\sqrt{-\Delta J_{\max}/2} u_{i,q}$, and $t_p \sim \gamma_m w_q/J_{\max}$. 
The collective dephasing term $\bra{\psi}\mathscr{D}[S]\ket{\psi}$ 
can be written as
\begin{eqnarray}
&&\bra{\psi}\mathscr{D}[S]\ket{\psi}=\sum_i \frac{g_i^2}{\Delta^2}(1-\bra{\psi} \sigma_i^z\ket{\psi}^2)
\nonumber \\
&& +\sum_{i\neq j} \frac{g_i g_j}{\Delta^2}(\bra{\psi} \sigma_i^z \sigma_j^z \ket{\psi}\!-\!\bra{\psi} \sigma_i^z\ket{\psi}\bra{\psi} \sigma_j^z\ket{\psi}) \nonumber.
\end{eqnarray}
The scaling of $\bra{\psi}\mathscr{D}[S]\ket{\psi}$ is in general nontrivial as it depends on the many-body structure of $\ket{\psi}$. 
However, our numerical results can be explained by considering that the first term dominates over the second term. 
This assumption is in particular valid around the initial and final times of the QAOA evolution when $\ket{\psi}$ is approximately a product state.
Considering then the worst case scenario $\bra{\psi} \sigma_i^z\ket{\psi} \approx 0$, and using $\sum_i |u_{i,q}|^2=1$, we indeed obtain the estimate Eq.~\eqref{eq:error_rethermalization} for the accumulated error for the total QAOA evolution $t_g\to t_\mathrm{run}$. Note that for other types of multi-qubit evolutions than QAOA, we cannot exclude the possibility that the second term plays a role and changes the error scaling.

\subsection{Cooperativity Parameter}

In this section we show that the two-qubit error can be expressed
in terms of a single cooperativity parameter $C$. 
Here, for simplicity we first consider a single resonator mode of frequency $\omega_{1}$, 
as could be realized based on parametric modulation of the qubit-resonator coupling \cite{harvey18SM, Royer2017SM}, 
with the replacement $\omega_1 \rightarrow \Delta$. 

\textit{Single mode setting.}---Following the main text, we consider
two error sources: (i) dephasing of the qubits on a timescale $\sim T_{2}$
and (ii) rethermalization of the resonator mode an with an effective
decay rate $\sim\kappa\bar{n}_{\mathrm{th}}$, with $\kappa=\omega_{1}/Q$.
The gate time is given by $t_{\mathrm{coh}}=\pi/4J_{12}$, with $J_{12}=g^{2}/\omega_{1}$
(we have set $g_{i}=g$ for simplicity). As shown above, both analytically
and numerically, the dephasing induced error can be expressed as $\xi_{\gamma}=\alpha_{\gamma}\gamma_{\phi}/J_{12}$,
with the pre-factor $\alpha_{\gamma}=N\pi/8$. The rethermalization-induced
error can be written as $\xi_{\kappa}=\alpha_{\kappa}\left(\kappa/\omega_{1}\right)\bar{n}_{\mathrm{th}}$,
as follows from multiplying the effective dephasing rate $\Gamma_{\mathrm{eff}}\sim\kappa\bar{n}_{\mathrm{th}}\left(g/\omega_{1}\right)^{2}$
with the gate time time $t_{\mathrm{coh}}=\pi/4J_{12}\approx\omega_{1}/g^{2}$
\cite{schuetz17SM}; the pre-factor $\alpha_{\kappa}$ can be obtained
numerically as $\alpha_{\kappa}\approx3$. In the small error regime,
we can add up these two errors independently and arrive at the total
error 
\begin{equation}
\xi=\alpha_{\kappa}\frac{k_{B}T}{Q\omega_{1}}+\alpha_{\gamma}\frac{\gamma_{\phi}\omega_{1}}{g^{2}},
\end{equation}
where we have used $\kappa\bar{n}_{\mathrm{th}}\approx k_{B}T/Q$.
For fixed spin-photon coupling $g$, this general expression for $\xi$
can be optimized with respect to the frequency $\omega_{1}$. The
optimal frequency $\omega_{1}^{\star}$ is given as 
\begin{equation}
\omega_{1}^{\star}=\sqrt{\frac{\alpha_{\kappa}}{\alpha_{\gamma}}\frac{k_{B}Tg^{2}}{Q\gamma_{\phi}}}.
\end{equation}
For faster dephasing $\sim\gamma_{\phi}$, the optimal value of $\omega_{1}^{\star}$
decreases, to allow for a faster gate (since $J\sim1/\omega_{1}$),
while $\omega_{1}^{\star}$ increases with larger rethermalization
$\kappa\bar{n}_{\mathrm{th}}\approx k_{B}T/Q$, because the thermal
occupation will be smaller. For this optimized value of $\omega_{1}^{\star}$,
the error $\xi$ simplifies to 
\begin{equation}
\xi=\frac{2\sqrt{\alpha_{\kappa}\alpha_{\gamma}}}{\sqrt{C}}\sim\frac{1}{\sqrt{C}},
\end{equation}
where we have introduced the cooperativity parameter as 
\begin{equation}
C=\frac{g^{2}}{\gamma_{\phi}\kappa\bar{n}_{\mathrm{th}}}\approx\frac{g^{2}Q}{\gamma_{\phi}k_{B}T}.
\end{equation}
In essence, the parameter $C$ compares the coherent coupling $g$
with the geometric mean of the decoherence rates, given by $\gamma_{\phi}$
and $\kappa\bar{n}_{\mathrm{th}}$, respectively. 
Taking (for example)
$g/2\pi\approx10\mathrm{MHz}$, $T\approx1\mathrm{K}$, $Q\sim10^{5}$
and $\gamma_{\phi}/2\pi\approx0.1\mathrm{kHz}-0.1\mathrm{MHz}$ (corresponding
to $T_{2}\approx10\mu\mathrm{s}-10\mathrm{ms}$) \cite{schuetz17SM}, we obtain a cooperativity
in the range $C\approx5\times10^{3}$ up to $C\approx5\times10^{6}$,
yielding an overall two-qubit error $\xi$ in the range $\xi\approx\left(0.1-4.3\right)\%$.
For comparison, for the implementation of the QAOA protocol the decoherence error $\xi$ is amplified by both 
(i) the circuit depth $M$ and (ii) the larger number of qubits $N$, by a factor $\sim M N^{3/2}$. 
This increase can be compensated when using optimized parameters, say 
$g/2\pi\approx100\mathrm{MHz}$, $T\approx100\mathrm{mK}$, and $Q\sim10^{6}$.

\section{Implementation with superconducting qubits}

In this section, we propose an implementation of our model with superconducting qubits. 
Our approach is based on Ref.~\cite{Didier2015}, which we extend to two qubits and to the multi-mode scenario.

\begin{figure}
\includegraphics[width=\columnwidth]{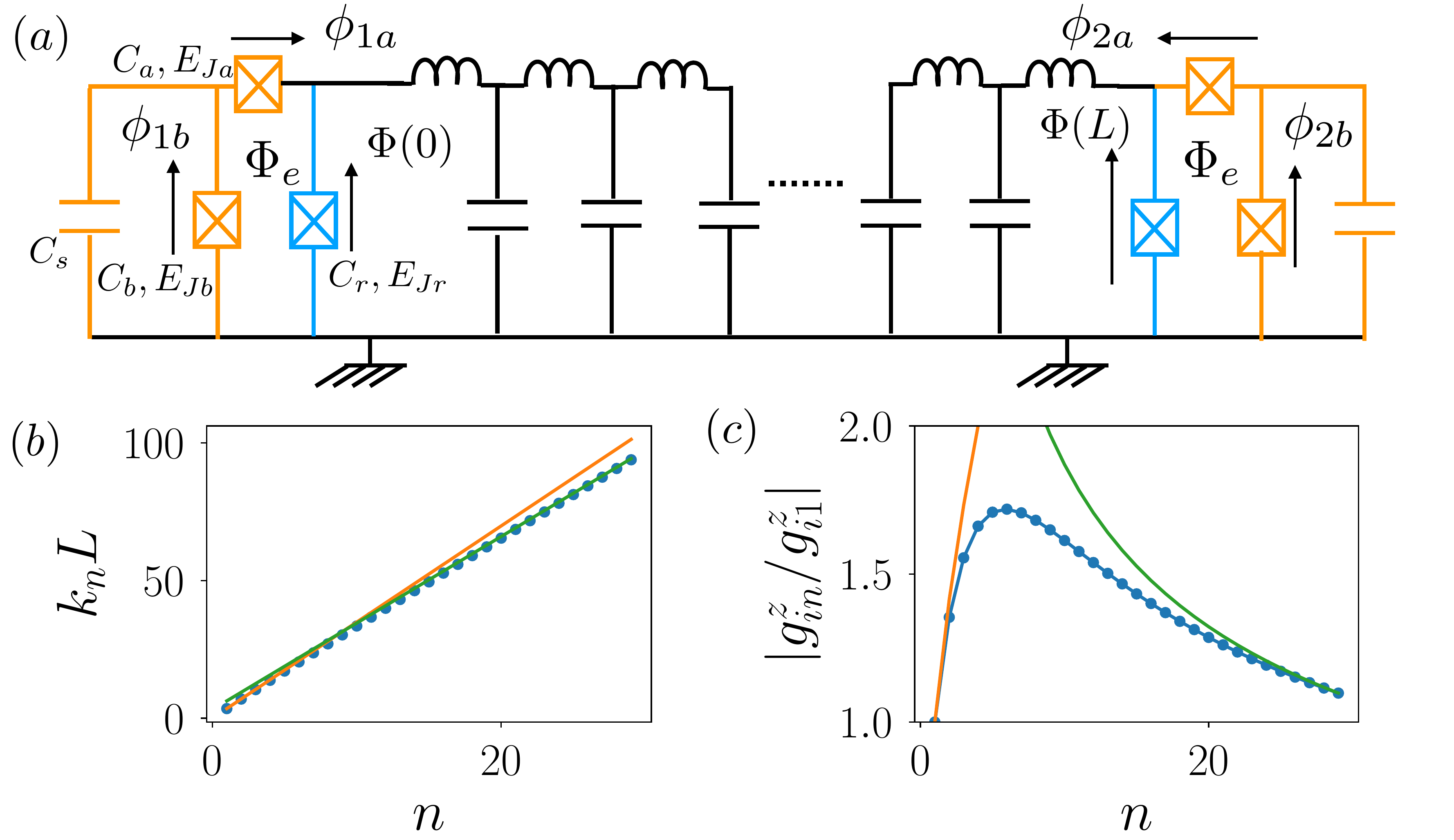}
\caption{{\it Implementation of longitudinal couplings with two transmon qubits.} (a) Circuit representation of our model with the two qubits placed at the two edges of a transmission line. The connecting inductances are shown in blue. (b) Dispersion relation and (c) and longitudinal couplings for $a_1=0.05L$. The asymptotic expressions (orange and green lines) are described in the text.\label{fig:setupSC}}
\end{figure}

\subsection{Setup}
The setup we have in mind is shown in Fig.~\ref{fig:setupSC} with two transmon qubits (depicted in orange) placed at the two edges of the transmission line.
Here, the connecting Josephson junctions (shown as blue)  create a phase drop in the transmission line and will lead to the desired longitudinal coupling . In the following, we show how to write the spin-Boson Hamiltonian describing this implementation and how to connect it to the model presented in the main text.

\subsection{Total Lagrangian}

Following the quantization procedure~\cite{Blais2004,Vool2017}, we write the total Lagrangian as 
\begin{eqnarray}
\mathcal{L} &=& \int_0^L dx \left( \frac{c \dot \Phi^2}{2} - \frac{(\partial_x \Phi)^2}{2\ell} \right) \nonumber \\
&+& \sum_{i=1,2} \Big[ E_{Jr} \cos \left( \frac{\Phi(x_i)}{\phi_0}\right) +   C_{r} \frac{\dot \Phi(x_i)^2}{2} \nonumber \\
&+& \left(\frac{C_s}{2} \dot\phi_{ib}^2 + \frac{C_a}{2} \dot\phi_{ia} ^2+ \frac{C_b}{2} \dot\phi_{ib}^2\right) \nonumber \\
&+& E_{Ja} \cos \left(\frac{\phi_{ia}}{\phi_0}\right)+ E_{Jb}\cos \left(\frac{\phi_{ib}}{\phi_0}\right)\Big], 
\end{eqnarray}
with $x_1=0,x_2=L$, $\phi_0=\hbar/(2e)$, and $c$ (resp. $\ell$) the capacitance (inductance) per unit length of the transmission line. Flux quantization in the transmon loops leads to the identities
\begin{eqnarray}
\phi_{ia}+\phi_{ib} &=& \Phi_e+ \Phi(x_i)\equiv \delta_i  \nonumber,
\end{eqnarray}
with $\Phi_e$ an applied external magnetic flux. Writing $\phi_{ia,ib}=\delta_i/2\mp\phi_i$, $\phi_i=(\phi_{ib}-\phi_{ia})/2$, we obtain
\begin{eqnarray}
\mathcal{L} &=& \int_0^L dx \left( \frac{c \dot \Phi^2}{2} - \frac{(\partial_x \Phi)^2}{2\ell} \right) \nonumber \\
&+& \sum_{i=1,2} \Big[ E_{Jr} \cos \left( \frac{\Phi(x_i)}{\phi_0}\right) +   C_{r} \frac{\dot \Phi(x_i)^2}{2} \nonumber \\
&+& \frac{C_s+C_b}{2} \left(\frac{\dot\Phi(x_i)}{2}+\dot\phi_i\right)^2 + \frac{C_a}{2} \left(\frac{\dot\Phi(x_i)}{2}-\dot\phi_i\right)^2 \nonumber \\
&+& E_{J}\cos \left(\frac{\delta_i}{2\phi_0}\right) \cos \left(\frac{\phi_i}{\phi_0}\right)\Big], 
\end{eqnarray}
where we assumed identical junction energies \mbox{$E_{Ja}=E_{Jb}\equiv E_J/2$, and $\dot \Phi_e=0$}. 
We now linearize in first order in $\Phi(0),\Phi(L)\ll \phi_0$ the cosine term $\propto \cos(\delta_i/\phi_0)$. This allows to write the Lagrangian as
\begin{equation}
\mathcal{L}=\mathcal{L}_0 +\mathcal{L}_\mathrm{int}, 
\end{equation}
representing respectively the transmission line and transmon qubits, and the coupling terms between them
\begin{eqnarray}
\mathcal{L}_0 &=& \int_0^L dx \left( \frac{c \dot \Phi^2}{2} - \frac{(\partial_x \Phi)^2}{2\ell} \right)
\nonumber \\
&+&\sum_i \frac{\mathcal{C}}{2} \dot \Phi(x_i)^2 - \frac{\Phi(x_i)^2}{2L_r}
\nonumber \\
&+& \frac{C_T}{2}\dot\phi_i^2 + E_J \cos\left (\frac{\Phi_e}{2\phi_0} \right)\cos\left(\frac{\phi_i}{\phi_0}\right) 
\nonumber \\
\mathcal{L}_\mathrm{int} &=& - \sum_i E_J \sin\left (\frac{\Phi_e}{2\phi_0}\right) \left(  \frac{\Phi(x_i)}{2\phi_0} \right)\cos\left(\frac{\phi_i}{\phi_0}\right) 
\nonumber \\
&+& \sum_i \frac{C'}{2} \dot \Phi(x_i) \dot \phi_i,
\end{eqnarray}
with $\mathcal{C}=C_{r}+C_T/4$, $L_r=\phi_0^2/E_{Jr}$ the Josephson inductance, $C_T = C_a+C_b+C_s$, and $C'=(C_s+C_b-C_a)/2$. It is important to note that the capacitance $\mathcal{C}$ and inductance $L_r$ act as boundary conditions for the transmission line and thus control the corresponding mode structure~\cite{Bourassa2009,Malekakhlagh2017,Gely2017}. Also, the interaction Lagrangian $\mathcal{L}_\mathrm{int}$ consists of two terms, representing respectively longitudinal and transverse couplings (see below) of the transmon qubits to the transmission line.


\subsection{Mode structure of the transmission line.}
In order to map our superconducting qubit implementation to the model presented in the main text, we diagonalize the transmission line contribution of the Lagrangian $\mathcal{L}_0$ (see also for instance Ref.~\cite{Malekakhlagh2017}) to obtain a basis of photon modes. To do so, we write the Euler-Lagrange equations for $\Phi(x,t)=u_n(x)e^{-i\omega_nt}$
\begin{eqnarray}
\partial_x^2 u_n(x)+v_p^2  \omega_n^2 u_n(x)=0,
\end{eqnarray}
with $v_p= 1/\sqrt{\ell c}$ the speed of light in the transmission line. Without loss of generality, we can write the mode functions as sine waves
\begin{eqnarray}
u_n(x)&=& \sqrt{\frac{2}{L}} \sin (k_nx-\theta_n),
\end{eqnarray}
with the dispersion relation $\omega_n=k_n v_p$, $\theta_n$ a real number, and with boundary conditions
\begin{eqnarray}
u'_n(x_i) = (-1)^i\left (\frac{1}{a_1}+k_n^2a_2 \right) u_n(x_i). 
\end{eqnarray}
Here $a_1=L_r/\ell$, and $a_2=\mathcal{C}/c$ are two lengths, representing the effective spatial extent of the transmission line-qubit coupling (see below).  We can finally rewrite the above equation in the form of the two coupled transcendental equations
\begin{eqnarray}
k_na_1&=&  (1+a_1a_2 k_n^2) \tan(\theta_n)\label{eq:trans1} \\
k_n &=& \frac{n\pi+2\theta_n}{L} \quad (n>1) .
\label{eq:trans2}
\end{eqnarray}
In the general case, the equations are solved numerically, and we discuss two asymptotic regimes below.
Writing $\Phi(t)=\sum_n \Phi_n(t) u_n(x)$, we can finally write
\begin{eqnarray}
\mathcal{L}_0&=&\frac{c}{2} \sum_n (\dot\Phi_n^2-\omega_n^2\Phi_n^2), 
\nonumber \\
 &+& \sum_i   \frac{C_T}{2}\dot\phi_i^2 + E_J \cos\left (\frac{\Phi_e}{2\phi_0} \right)\cos\left(\frac{\phi_i}{\phi_0}\right) 
 \nonumber \\
\mathcal{L}_\mathrm{int} &=&-E_J \sin\left (\frac{\Phi_e}{2\phi_0}\right) \sum_{n,i} u_n(x_i) \left(  \frac{\Phi_n}{2\phi_0} \right)\cos\left(\frac{\phi_i}{\phi_0}\right) 
\nonumber \\
&+& \frac{C'}{2}  \sum_{n,i} u_n(x_i) \dot \Phi_n \dot \phi_i,
\end{eqnarray}
where we assumed the functions $u_n(x)$ to be normalized (valid in limit $\theta_n \ll k_n L$).

\subsection{Hamiltonian description}

We can now perform a Legendre transformation, writing the charge degrees of freedom as
\begin{eqnarray}
q_n &=& \frac{\partial\mathcal{L}}{\partial \dot \Phi_n} = c \dot \Phi_n+\sum_i \frac{C'}{2}u_n(x_i) \dot \phi_i
\nonumber \\
q_i &=& \frac{\partial\mathcal{L}}{\partial \dot \phi_i} = C_T \dot \phi_i +\sum_n \frac{C'}{2}u_n(x_i) \dot \Phi_n.
\end{eqnarray}
In first order in $\mathcal{L}_\mathrm{int}/\mathcal{L}_0$, i.e assuming the capacitive energy of the coupling term ($\propto C'$) can be treated perturbatively, we obtain
\begin{eqnarray}
\dot \Phi_n&=& \frac{q_n }{c}-\sum_i \frac{q_i C' u_n(x_i) }{2cC_T}
\nonumber \\
\dot \phi_i &=&  \frac{q_i }{C_T}-\sum_n \frac{q_n C' u_n(x_i) }{2cC_T},
\end{eqnarray}
and thus \mbox{$H=\sum_n q_n \dot \Phi_n +\sum_i q_i \dot \phi_i-\mathcal{L}=H_0+H_\mathrm{int}$}
\begin{eqnarray}
H_0 &=& \sum_n \left( \frac{q_n^2}{2c}+\frac{\omega_n^2c\Phi_n^2}{2} \right)
\nonumber \\
&+& \sum_i \frac{q_i^2}{2C_T} -E_J \cos\left(\frac{\Phi_e}{2\phi_0}\right)  \cos\left(\frac{\phi_i}{\phi_0}\right)
\nonumber \\
H_\mathrm{int}&=&- \frac{C'}{2C_Tc} \sum_{i,n} u_n(x_i) q_n q_i
\nonumber \\
&+& \frac{E_J}{2\phi_0}  \sin\left(\frac{\Phi_e}{2\phi_0}\right)   \sum_{i,n}  u_n(x_i) \Phi_n  \cos\left(\frac{\phi_i}{\phi_0}\right).\nonumber
\end{eqnarray}
Assuming for simplicity the transmon to be in the linear regime~\footnote{At the next order, we obtain the qubit nonlinearity.}, we can rewrite the first term as
\begin{eqnarray}
H_0  &=& \sum_n \left(\frac{q_n^2}{2c}+\frac{\omega_n^2c\Phi_n^2}{2} \right)+ \sum_i \left(\frac{q_i^2}{2C_T} +\frac{\omega_z^2C_T\phi_i^2}{2}\right), \nonumber
\end{eqnarray}
with $\omega_z =  1/\sqrt{L_T C_T}$ the qubit frequency, $L_T=\phi_0^2/E_J(\Phi_e)$, and $E_J(\Phi_e)=E_J\cos\left(\frac{\Phi_e}{2\phi_0}\right)$, 
which we can diagonalize in terms of harmonic oscillator operators describing the transmon and transmission line excitations
\begin{eqnarray}
\Phi_n &=& \sqrt{\frac{\hbar }{2c\omega_n}}(a_n+a_n^\dagger),
\quad
q_n =  \sqrt{\frac{\hbar c\omega_n}{2}}i(a_n^\dagger-a_n)
\nonumber \\
\phi_i &=& \sqrt{\frac{\hbar }{2C_T\omega_z}}(a_i+a^\dagger_i),
\quad
q_i =  \sqrt{\frac{\hbar C_T\omega_z}{2}}i(a^\dagger_i-a_i) \nonumber,
\end{eqnarray}
to obtain
\begin{eqnarray}
H_0  &=& \sum_n \hbar \omega_n a^\dagger_n a_n + \hbar \omega_z \sum_i  a^\dagger_i a_i.
\end{eqnarray}
Finally, in terms of these eigenmodes, the coupling Hamiltonian reads in the $\{0_i,1_i\}$ subspace of the qubits
\begin{eqnarray}
H_\mathrm{int}   &=&\hbar\sum_{i,n} \Omega_{i,n} (a_n^\dagger +a_n)+  \hbar\sum_{i,n}  g^z_{in} \sigma_i^z (a_n^\dagger +a_n),
\nonumber \\
&+&\hbar \sum_{i,n}  g^y_{in}  (a_i^\dagger - a_i)(a_n^\dagger - a_n)\label{eq:Hint}, 
\end{eqnarray}
with the couplings frequencies
\begin{eqnarray} 
\Omega_{in} &=&  \frac{E_J}{2\phi_0}  \sin\left(\frac{\Phi_e}{2\phi_0}\right)  \sqrt{\frac{1}{2\hbar c\omega_n}}u_n(x_i) \frac{A_{1}+A_{0}}{2}
\nonumber \\ 
g^z_{in} &=& \frac{E_J}{2\phi_0}  \sin\left(\frac{\Phi_e}{2\phi_0}\right)  \sqrt{\frac{1}{2\hbar c\omega_n}}u_n(x_i) \frac{A_{1}-A_{0}}{2}
\nonumber \\ 
g^y_{in} &=&\frac{C'}{2C_Tc} \sqrt{\frac{C_T\omega_z}{2}}  \sqrt{\frac{ c\omega_n}{2}}u_n(x_i),
\nonumber \\ 
\end{eqnarray}
and matrix elements $A_{s_i}=\bra{s_i}\cos(\phi_i/\phi_0)\ket{s_i}$ for the qubit operators. The first term in Eq.~\eqref{eq:Hint} is a driving term creating photons in the transmission line due to the presence of the external flux $\Phi_e$, and which is absent in our model Eq.~(1) of the MT. Note however that this term can be eliminated using displaced bosonic operators $a_n\to a_n +\Omega_n/\omega_n$.
The second term represents the desired longitudinal interactions, and scales with the qubit junction energy $E_J$ and can be tuned by the external flux $\Phi_e$~\cite{Didier2015}. We discuss the multimode structure and origin of the frequency cutoff below. Finally, the last term is a transverse coupling whose strength is controlled by the different capacitances of the qubits. Interestingly, we can eliminate this term by setting $C'=0$, i.e. $C_s=C_a+C_b$.

\subsection{Numerical results and asymptotic expressions}

To conclude our implementation, we analyse the form of dispersion relation of the transmission line, and the scaling of the coupling term $g_{in}^z$ with respect to the mode number $n$, assuming for simplicity $C'=0$ (no transverse coupling) and $a_2/a_1 \approx 0$ (the frequency cutoff is only set by the inductance $L_r$ of the connecting junction). 

The dispersion relation, calculated by numerically solving Eqs.~\eqref{eq:trans1},\eqref{eq:trans2} for $a_1=0.05L$ is shown in Fig.~\ref{fig:setupSC}(b), and is close to being linear. We represent in panel (c) the corresponding coupling strengths $g_{in}^z$.
At small spatial frequencies $k_n a_1\ll1 $, we can linearize Eqs.~\eqref{eq:trans1},\eqref{eq:trans2} and obtain asympotic expressions for $k_n\approx \pi  n/(L-2 a_1)$, and $g^z_{in} \propto \sqrt{n}$,
Similarly, at high frequencies, $k_n a_1\gg1$, we have instead
$k_n\approx (n+1)\pi/L$, $g_{in} \propto  1/\sqrt{n}$. 
These asymptotic expressions for $k_n$ and $g_{in}$ are shown as blue and orange line respectively. 
Note that such quasi linear dispersion relation and form of the coupling $g_{in}^z$ have also been shown in the case of transverse couplings~\cite{Malekakhlagh2017,Gely2017}. Also, the scalings with $n$ in the low and high-frequency regime of $g_{in}^z$ match the phenomenological expression used in the main text.

Note that our model can be generalized to the $N$ qubits scenario. This would require however a complete numerical analysis to compute the mode structure of the transmission line, obtain the corresponding Hamiltonian, and assess the magnitude of longitudinal and possible residual transverse couplings. Approaches based on capacitive couplings of asymmetric flux qubits to the transmission line~\cite{Billangeon2015} represent another interesting option, where the frequency cutoff is determined by the coupling capacitance~\cite{Malekakhlagh2017,Gely2017}.


\subsection{Typical numbers}
We conclude this section by giving relevant numbers and error estimates for a SC implementation of our model. 
The estimated gate time between qubits induced by longitudinal couplings is of the order of $\sim 50$ ns, corresponding to couplings $g/(2\pi) \approx 60 $ MHz~\cite{Royer2017SM}. 
For concreteness, we consider a coherence time $T_2\approx 100\ \mu$s~\cite{Gambetta2017SM}, a loss rate $\kappa/(2\pi)=0.05 $ MHz~\cite{Royer2017SM} and a thermal population of $\bar{n}_\mathrm{th}(\omega_0)=3$. 
These numbers correspond to a cooperativity $C\approx 6\times10^6$, which translates to a total QAOA error of about $\sim 8\%$ for $N=12$ qubits and $M=5$ QAOA cycles.
For the same set of parameters, as indicated in the main text, the backbone of our QAOA implementation, namely the two-qubit hot gate (with $N=2$ and $M=1$) and the spin engineering recipe (where $M=1$) could be demonstrated with considerably smaller errors.

\section{Implementation with Quantum Dots}



In this Appendix we provide background material for the implementation of our 
theoretical scheme using quantum dots coupled to transmission line resonators. 
First, we derive the microscopic coupling between a quantum dot and a multi-mode microwave cavity. 
If one restricts oneself to the lowest dot orbital and a single resonator mode we recover standard expressions used in the literature. 
Specifically, we then discuss quantum dot charge qubits and singlet-triplet qubits.

\subsection{Microscopic Dot-Resonator Coupling}

Following Refs.\cite{viennot16SM,cottet15SM}, microscopically the coupling
between a quantum dot, described by its electron density $\rho\left(\mathbf{r}\right)=\sum_{\sigma}\Psi_{\sigma}^{\dagger}\left(\mathbf{r}\right)\Psi_{\sigma}\left(\mathbf{r}\right)$,
to a microwave cavity with associated voltage fluctuations $\hat{V}\left(\mathbf{r}\right)=\sum_{n}\phi_{n}\left(\mathbf{r}\right)\left(a_{n}+a_{n}^{\dagger}\right)$
{[}for convenience we have assumed the voltage mode functions $\phi_{n}\left(\mathbf{r}\right)$
to be real{]} is given by 
\begin{eqnarray}
H_{I} & = & e\int d\mathbf{r}\hat{V}\left(\mathbf{r}\right)\rho\left(\mathbf{r}\right),\\
 & = & e\sum_{\sigma}\int d\mathbf{r}\hat{V}\left(\mathbf{r}\right)\Psi_{\sigma}^{\dagger}\left(\mathbf{r}\right)\Psi_{\sigma}\left(\mathbf{r}\right),
\end{eqnarray}
where $e$ is the electron's charge. Next, we express the field operator
$\Psi_{\sigma}\left(\mathbf{r}\right)$ in terms of the annihilation
operators associated with the dot orbitals $\nu_{i}$ of dot $i$
as 
\begin{equation}
\Psi_{\sigma}\left(\mathbf{r}\right)=\sum_{i,\nu_{i}}\varphi_{i\nu_{i}}\left(\mathbf{r}\right)c_{i\nu_{i},\sigma}.
\end{equation}
Here, the fermionic operator $c_{i\nu_{i},\sigma}(c_{i\nu_{i},\sigma}^{\dagger})$
annihilates (creates) an electron of spin $\sigma=\uparrow,\downarrow$
in the orbital $\nu_{i}$ of dot number $i$. We then arrive at 
\begin{equation}
H_{I}=\sum_{n}\sum_{i,\nu_{i},j,\nu_{j},\sigma}g_{n,i,\nu_{i},j,\nu_{j}}\left(a_{n}+a_{n}^{\dagger}\right)c_{i\nu_{i},\sigma}^{\dagger}c_{j\nu_{j},\sigma},
\end{equation}
where 
\begin{equation}
g_{n,i,\nu_{i},j,\nu_{j}}=e\int d\mathbf{r}\phi_{n}\left(\mathbf{r}\right)\varphi_{i\nu_{i}}^{*}\left(\mathbf{r}\right)\varphi_{j\nu_{j}}\left(\mathbf{r}\right).
\end{equation}
For standard geometries---compare for example Refs.~\cite{childress04SM, frey12SM, beaudoin16SM}---where (for example) only one dot
out of a larger double quantum dot is exposed to the resonator's voltage
the mode function $\phi_{n}\left(\mathbf{r}\right)$ overlaps only
with one specific dot, say the right dot (labeled by $R$). Therefore,
to obtain a non-zero expression for the coupling $g_{n,i,\nu_{i},j,\nu_{j}}$,
we can fix one of the indices as $i=R$ or $j=R$ (for $i=L,R$ in
a DQD setting). Following Ref.\cite{cottet15SM}, we neglect photon-induced
orbital tunneling terms because of small wavefunction overlap and
focus on the dominant coupling term where $i=j=R$; further suppression
of photon-induced tunneling terms can be achieved by carefully avoiding
resonances between the resonator frequencies $\omega_{n}$ and the
(tunable) transition energies $\Delta_{i\nu_{i},j\nu_{j}}$. In this
case the dot-resonator coupling reduces to 
\begin{equation}
H_{I}=\sum_{n}\sum_{\nu,\nu',\sigma}g_{n\nu\nu'}\left(a_{n}+a_{n}^{\dagger}\right)c_{R\nu\sigma}^{\dagger}c_{R\nu'\sigma},\label{eq:dot-resonator-coupling}
\end{equation}
with 
\begin{equation}
g_{n\nu\nu'}=e\int d\mathbf{r}\phi_{n}\left(\mathbf{r}\right)\varphi_{R\nu}^{*}\left(\mathbf{r}\right)\varphi_{R\nu'}\left(\mathbf{r}\right).
\end{equation}
Next, as detailed in Ref.\cite{cottet15SM}, we drop photon-induced
orbital tunneling terms within one dot where $\nu\neq\nu'$, because
of negligible wavefunction overlap. 
Within this approximation, we
recover the standard form for capacitive dot-resonator coupling \cite{childress04SM}
\begin{equation}
H_{I}=\sum_{n,\nu}g_{n\nu}\left(a_{n}+a_{n}^{\dagger}\right)\otimes\hat{n}_{R\nu}, \label{eq:H-int-QD-multi-electron}
\end{equation}
with $\hat{n}_{R\nu}=\sum_{\sigma}c_{R\nu\sigma}^{\dagger}c_{R\nu\sigma}$
and 
\begin{equation}
g_{n\nu}=e\int d\mathbf{r}\phi_{n}\left(\mathbf{r}\right)\left|\varphi_{R\nu}\left(\mathbf{r}\right)\right|^{2}.
\end{equation}
If one restricts oneself to the lowest electronic orbital $\nu=0$
and a single resonator mode $n$ we recover standard expressions as
used for example in Refs.\cite{taylor06SM,beaudoin16SM,childress04SM,viennot16SM,cottet15SM}.
Within this description the resonator's voltage fluctuations amount
to \textit{fluctuations of the dot's chemical potential} \cite{harvey18SM},
with a coupling strength approximately given by $g_{n0}\approx e\phi_{n}\left(\mathbf{r}_{\mathrm{dot}}\right)$
where $\mathbf{r}_{\mathrm{dot}}$ refers to the center of the wavefunction
$\varphi_{R0}\left(\mathbf{r}\right)$; this treatment amounts to
the dipole-like approximation in quantum optics where the quantum
dot is considered as point-like on the relevant lengthscale set by
the wavelength associated with the mode-function $\phi_{n}\left(\mathbf{r}\right)$.
This can be done for long-wavelength resonator modes, but eventually
the coupling $g_{n\nu}$ will average out for sufficiently large mode
number $n$ because of rapid oscillations of the associated mode function
$\phi_{n}\left(\mathbf{r}\right)$, as is well known also from coupling
of quantum dots to phonons \cite{hanson07SM}. 
While this is the case for very large $n$ only, our microscopic form of the 
spin-resonator coupling avoids unphysical divergencies and ensures a finite 
spin-spin coupling parameters $J_{ij}$, as shown in detail in Sec. \ref{Effective-spin-spin-interactions}. 

In the low temperature regime of interest (where $k_{B}T\ll\Delta_{\mathrm{orb}}$)
we can restrict ourselves to the lowest electronic orbital $\nu=0$.
In an effectively one-dimensional problem as considered here, with
a qubit localized at $x_{i}$, we can then express the \textit{charge-based}
coupling between qubit $i$ and mode $n$ as 
\begin{equation}
g_{n}=g_{0}\sqrt{n}\int dx\cos\left(k_{n}x\right)f\left(x-x_{i}\right). \label{eq:QD-microscopic-coupling}
\end{equation}
Here, we have set $\phi_{n}\left(x\right)=\phi_{n}\cos\left(k_{n}x\right)$,
$f\left(x-x_{i}\right)=\int dydz\left|\varphi_{R0}\left(\mathbf{r}\right)\right|^{2}$
and $g_{0}\sqrt{n}=e\phi_{n}$, with the single photon voltage fluctuation
amplitude $\phi_{n}=\alpha V_{n}\sim\sqrt{n}$; here, the amplitudes
$\phi_{n}$ account for the potential fluctuations felt by the quantum
dot via the lever arm $\alpha$. 
This expression matches the one used in the main text where $g_{i,n}$ refers to qubit $i=1,\dots,N$, with individual amplitudes $g_{0} \rightarrow g_{i}$.


\textit{Quantum dot orbital transitions.---}Typically, for gate-defined quantum dots the single-particle orbital
level $\Delta_{\mathrm{orb}}$ spacing amounts to $\Delta_{\mathrm{orb}}\sim1\mathrm{meV}$
\cite{cerletti05SM}. This energy scale is much larger than the thermal
energy $k_{B}T$ for temperatures as high as $T=1\mathrm{K}$ which
corresponds to $k_{B}\times1\mathrm{K}\approx8.6\times10^{-2}\mathrm{meV}$.
For comparatively high temperatures in the range $T\approx\left(1-4\right)\mathrm{K}$,
the thermal occupation of photons with energy $\hbar\omega=\Delta_{\mathrm{orb}}$
is $\bar{n}_{\mathrm{th}}\left(\Delta_{\mathrm{orb}}\right)\approx10^{-5}-6\times10^{-2}$.
Therefore, the overwhelming majority of quantum dot experiments (which
typically operate at dilution fridge temperatures where $T\approx\left(10-100\right)\mathrm{mK}\ll1\mathrm{K}$)
can be described by restricting oneself to the orbital ground-state
subspace. Along the same lines, we will restrict ourselves to the
lowest orbital levels, since $\Delta_{\mathrm{orb}}$ is much larger
than all relevant energy scales in our problem
\footnote{The temperature requirements may be more stringent (for example) in silicon samples, 
where the valley splitting has to be taken into account, which tends to be smaller than the orbital splitting.
In Si/SiGe quantum dots, the valley splitting is usually not larger than $\sim 100\mu \mathrm{eV}$, 
while it is larger in Si/SiO2 dots, in the range of hundreds of $\mu\mathrm{eV}$ up to $\sim 1 \mathrm{meV}$.}.

\subsection{Quantum Dot Charge Qubits}

\begin{figure}
\includegraphics[width=1\columnwidth]{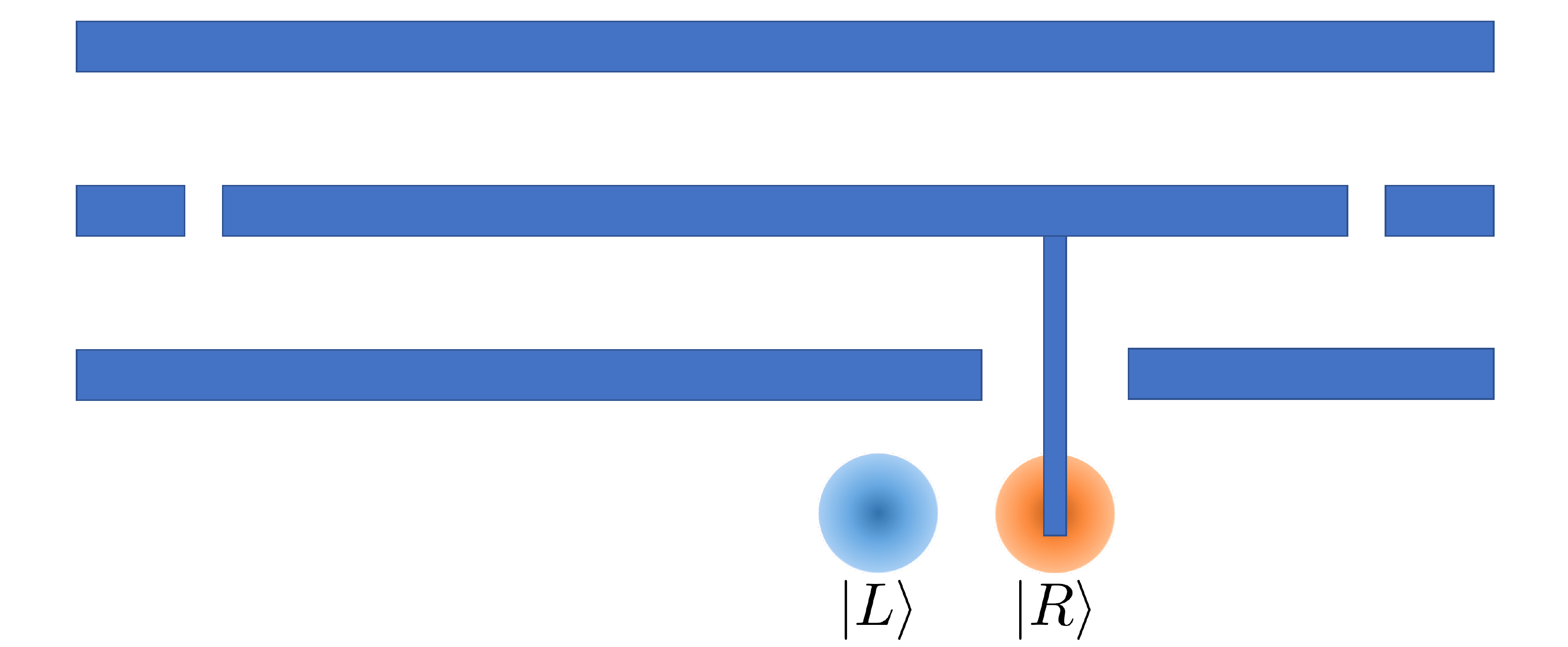}

\caption{\label{fig:implementation-DQD}
Schematic illustration of a DQD coupled to a transmission line. 
The charge-resonator coupling derives from the capacitive coupling to the right (or left) dot only. 
Here, the right dot is taken to be sensitive to the zero-point voltage of a coplanar-waveguide resonator via a capacitive
finger with so-called lever arm $\alpha$, which couples microwave photons in the resonator to the orbital degree-of-freedom of the electron \cite{beaudoin16SM,frey12SM}. 
Coupling between the resonator mode and the electron's spin can be achieved by making use of various mechanisms which hybridize spin and charge degrees of freedom, 
as provided by spin-orbit interaction or inhomogeneous magnetic fields \cite{beaudoin16SM, hu12SM, trif08SM, viennot15SM}.}
\end{figure}

In this subsection we briefly discuss a potential quantum-dot based physical implementation
of our scheme, closely following Ref.\cite{childress04SM};
for a schematic illustration compare Fig.\ref{fig:implementation-DQD}.
Consider a DQD in the \textit{single-electron} regime; below we will
separately consider double dots in the two-electron regime. The electron
can occupy the left $\left|L\right\rangle $ or right orbital $\left|R\right\rangle $,
respectively. The right dot is capacitively coupled to the resonator.
In this scenario, we can project the general result given in Eq.\eqref{eq:H-int-QD-multi-electron} onto the single-electron regime. 
When restricting ourselves to the lowest QD orbital as discused above, 
the interaction between the DQD and the transmission line can then be written as 
\begin{equation}
H_{I}=\sum_{n}g_{n}(a_{n} + a_{n}^{\dagger})\otimes\left|R\right\rangle \left\langle R\right|, \label{eq:H-int-QD-single-electron}
\end{equation}
with the coupling $g_{n}$ given in Eq.\eqref{eq:QD-microscopic-coupling}; 
this coupling accounts for a frequency cut-off as set by the microscopic size given by $f(x)$ 
[compare the main text where $f(x)$ has been modeled by a simple box-function with spatial extent $a$]. 
Note that, when neglecting this cut-off, we recover standard results as presented for example in \cite{childress04SM}. 
To make this comparison concrete, we can rewrite $H_{I}$ as
$H_{I}=ev\hat{V}\otimes\left|R\right\rangle \left\langle R\right|,$
where $e$ is the electron's charge, $\hat{V}$ is the (quantized)
voltage on the resonator near the right dot, $v=C_{c}/\left(C_{c}+C_{d}\right)$,
and $C_{d}$ is the capacitance to ground of the right dot; $C_{c}$
is the capacitive coupling between the right dot and the resonator. 
Following Ref.\cite{childress04SM,taylor06SM}, the quantized voltage
at the end of the transmission line (with length $L$) can be written as 
$\hat{V}=\sum_{n}\sqrt{\hbar\omega_{n}/LC_{0}}\left(a_{n}+a_{n}^{\dagger}\right)$
where $C_{0}$ refers to the capacitance per unit length; the allowed
wavevectors can be written as $k_{n}=\left(n+1\right)\pi/L$ and the
corresponding frequencies read $\omega_{n}=k_{n}/C_{0}Z_{0}$, with
the characteristic impedance $Z_{0}$ \cite{childress04SM}. 
Using (in our notation) $\omega_{n}=n\pi c/L$ (with $n=1,2,\dots$), the amplitudes of the zero-point fluctuations
can be written as 
$V_{n}=\sqrt{\hbar n\pi c/L^{2}C_{0}}$.
Accordingly, since the individual couplings $g_{i,n}$ scale with
the zero-point fluctuations, we find $g_{i,n}\sim\sqrt{n}/L$, in
direct agreement with Ref.\cite{sundaresan15}. 
The zero-point fluctuations (and therefore the qubit resonator coupling) can be increased significantly
with the help of so-called high-impedance resonators, as demonstrated
(for example) in Refs.\cite{stockklauser17SM,samkharadze16SM}. 
The voltage along such a high-impedance resonator
is much larger than for a conventional $50\Omega$ resonator, with
the single-photon voltage at the resonator's antinode being $V_{1}=\sqrt{\hbar Z_{r}}\omega_{1}$
for the fundamental mode $\omega_{1}$ \cite{harvey18SM}.

Coming back to the dot-resonator coupling described by Eq.\eqref{eq:H-int-QD-single-electron}, 
it is instructive to express $H_{I}$ in terms of the (orbital) DQD eigenstates, defined as \cite{childress04SM}
\begin{eqnarray}
\left|+\right\rangle  & = & \sin\theta\left|L\right\rangle +\cos\theta\left|R\right\rangle ,\\
\left|-\right\rangle  & = & \cos\theta\left|L\right\rangle -\sin\theta\left|R\right\rangle ,
\end{eqnarray}
where $\tan\theta=-2t_{c}/\left(\omega_{q}+\epsilon\right)$; the effective
qubit splitting $\omega_{q}=\sqrt{4t_{c}^{2}+\epsilon^{2}}$ between
the eigenstates $\left|+\right\rangle $ and $\left|-\right\rangle $
can be tuned in situ via the (tunable) tunnel-coupling $t_{c}$ and/or
the detuning parameter $\epsilon$. Defining the Pauli spin matrices (in the orbital pseudo-spin space)
as $\sigma^{+}=\left|+\right\rangle \left\langle -\right|$ etc.,
the full Hamiltonian can be written as 
\begin{eqnarray}
H & = & \frac{\omega_{q}}{2}\sigma^{z}+\sum_{n}\omega_{n}a_{n}^{\dagger}a_{n}\nonumber \\
 &  & +\sum_{n}\left(g_{n}^{z}\sigma^{z}+g_{n}^{x}\sigma^{x}\right)\otimes\left(a_{n}+a_{n}^{\dagger}\right),
\end{eqnarray}
with $\omega_{n}=\left(n+1\right)\omega_{0}$. 
The splitting $\omega_{q}$ can be tuned via the dot parameters $t_c$ and $\epsilon$, respectively, while the coupling constants
$g_{m}^{x},g_{m}^{z}$ can be controlled via the dot parameters as
$g_{n}^{x} = g_{n} t_{c}/\omega_{q}$, and 
$g_{n}^{z} = g_{n} \epsilon / 2\omega_{q}$ \cite{childress04SM}.
Again, when disregarding cut-off effects, we recover results as presented (for example) in \cite{childress04SM}, 
where $g_{n} = g_{0} \sqrt{n}$, with the overall coupling strength $g_{0}/\omega_{0}=v\sqrt{2Z_{0}/R_{Q}}=\mathrm{const.}$,
with $Z_{0}$ being the characteristic impedance of the transmission
line and $R_{Q}=h/e^{2}$ referring to the resistance quantum \cite{childress04SM}. 

Typically, as done in Ref.\cite{childress04SM}, one proceeds by considering
the regime $\omega_{0}\approx\omega_{q}$ in which one can neglect
all but (for example) the fundamental mode (setting $a_{0}=a$), eventually leading
to the Jaynes-Cummings Hamiltonian within a rotating wave approximation
\begin{equation}
H\approx\frac{\omega_{q}}{2}\sigma^{z}+\omega_{0}a^{\dagger}a+g_{0}^{x}\left(a\sigma^{+}+a^{\dagger}\sigma^{-}\right).
\end{equation}
This limit has been studied experimentally in Ref.\cite{frey12SM},
with a charge(pseudo-spin)-resonator strength of several tens of MHz. 

Conversely, for $t_{c}=0$ we realize a model with purely \textit{longitudinal}
charge(pseudo-spin)-resonator coupling, that is 
\begin{equation}
H=\frac{\omega_{q}}{2}\sigma^{z}+\sum_{n}\omega_{n}a_{n}^{\dagger}a_{n}+\sum_{n}g_{n}^{z}\sigma^{z}\otimes\left(a_{n}+a_{n}^{\dagger}\right).
\end{equation}
By indexing $\sigma^{z} \rightarrow \sigma_{i}^{z}$, $\omega_{q} \rightarrow \omega_{i}$ and $g_{n}^{z} \rightarrow g_{i,n}$, this single-mode Hamiltonian can be generalized to the multi-qubit scenario, as considered in the main text. 

\subsection{Singlet-Triplet Qubits}

In this section we show how to parametrically modulate the spin resonator
coupling in the case of singlet-triplet qubits embedded in DQDs. To make our work self-contained we first closely follow Ref.\cite{harvey18SM}
for a single-mode analysis, and then generalize this idea to a multi-mode
setup.


\textit{Single-Mode Setting.---}We follow Ref.\cite{harvey18SM} to analyze the coupling between singlet-triplet
qubits and a high-impedance resonator. When neglecting higher orbitals
and other spin levels (within the lowest orbital) the Hamiltonian
associated with a double quantum dot in the two-electron regime can
be written as 
\begin{equation}
H_{q}=\frac{J\left(\epsilon\right)}{2}\sigma^{z},
\end{equation}
with the exchange-induced splitting $J\left(\epsilon\right)=\omega_{q}\left(\epsilon\right)$
between the two qubit states $\left\{ \left|T_{0}\right\rangle ,\left|S\right\rangle \right\} $.
The detuning parameter $\epsilon$ can be readily controlled classically,
but also coupled to the quantized voltage fluctuations associated
with the resonator. When writing the detuning as $\epsilon=\epsilon_{0}+\delta\epsilon$
we can expand the splitting $J\left(\epsilon\right)$ in a Taylor
series around the equilibrium $\epsilon_{0}$ as 
\begin{equation}
J\left(\epsilon\right)\approx J\left(\epsilon_{0}\right)+J'\left(\epsilon_{0}\right)\delta\epsilon+\frac{1}{2}J''\left(\epsilon_{0}\right)\delta\epsilon^{2}+\dots
\end{equation}
In the presence of both classical driving and quantum fluctuations,
we have 
\begin{eqnarray}
\delta\epsilon & = & ec_{g}V_{g}\left(t\right)+ec_{r}\hat{V},\\
 & = & \epsilon_{d}\cos\left(\omega_{d}t\right)+ec_{r}\hat{V},
\end{eqnarray}
where $e$ refers to the electron charge, and $c_{g}$ ($c_{r}$)
give the geometrical lever arms between the double dot and the RF
gate (resonator); the latter sets the shift in chemical potential of
the DQD caused by a voltage shift on those gates, respectively. 

For a \textit{single-mode} resonator, where $\hat{V}=V_{0}\left(a+a^{\dagger}\right)$,
the total Hamiltonian 
\begin{equation}
H=\omega_{0}a^{\dagger}a+\frac{J\left(\epsilon\right)}{2}\sigma^{z},
\end{equation}
can then be approximated as \begin{widetext} 
\begin{eqnarray}
H & = & \omega_{0}a^{\dagger}a+\frac{1}{2}\left[J\left(\epsilon_{0}\right)+\frac{1}{2}J''\left(\epsilon_{0}\right)\left(\frac{\epsilon_{d}^{2}}{2}+e^{2}c_{r}^{2}V_{0}^{2}\right)\right]\sigma^{z}\\
 &  & +\frac{1}{2}J'\left(\epsilon_{0}\right)\left[\epsilon_{d}\cos\left(\omega_{d}t\right)+ec_{r}V_{0}\left(a+a^{\dagger}\right)\right]\sigma^{z}\nonumber \\
 &  & +\frac{1}{4}J''\left(\epsilon_{0}\right)\left[ec_{r}V_{0}\epsilon_{d}\left(e^{i\omega_{d}t}+e^{-i\omega_{d}t}\right)\left(a+a^{\dagger}\right)+2e^{2}c_{r}^{2}V_{0}^{2}a^{\dagger}a+e^{2}c_{r}^{2}V_{0}^{2}\left(a^{2}+\mathrm{h.c.}\right)+\frac{\epsilon_{d}^{2}}{2}\cos\left(2\omega_{d}t\right)\right]\sigma^{z}.\nonumber 
\end{eqnarray}

\end{widetext}

In the absence of driving $\left(\epsilon_{d}=0\right)$ to leading
linear order we would obtain 
\begin{equation}
H\approx\omega_{0}a^{\dagger}a+\frac{\tilde{J}\left(\epsilon_{0}\right)}{2}\sigma^{z}+g_{1}\sigma^{z}\otimes\left(a+a^{\dagger}\right),
\end{equation}
with a renormalized qubit splitting $\tilde{J}\left(\epsilon_{0}\right)$
and 
\begin{equation}
g_{1}=\frac{1}{2}J'\left(\epsilon_{0}\right)ec_{r}V_{0}.
\end{equation}
The maximum coupling $g_{1}$ is largely set by the zero-point voltage
fluctuation amplitude $V_{0}$, while the coupling can be tuned by
choosing the operating point $J'\left(\epsilon_{0}\right)$ appropriatedly.
The effective dipole is turned off (on) if $J'\left(\epsilon_{0}\right)=0$
$\left(|J'\left(\epsilon_{0}\right)|>0\right)$. 

In the presence of driving $\left(\epsilon_{d}>0\right)$, however,
in an interaction picture with respect to $H_{0}=\omega_{d}a^{\dagger}a+\tilde{J}\left(\epsilon_{0}\right)\sigma^{z}/2$
one can approximately (within a RWA, where all fast-oscillating terms
are dropped) restrict oneself to \cite{harvey18SM} 
\begin{equation}
\tilde{H}\approx\Delta a^{\dagger}a+g_{2}\sigma^{z}\otimes\left(a+a^{\dagger}\right),
\end{equation}
with the detuning $\Delta=\omega_{0}-\omega_{d}$ and the coupling
\begin{equation}
g_{2}=\frac{1}{4}J''\left(\epsilon_{0}\right)ec_{r}V_{0}\epsilon_{d}.
\end{equation}
The coupling $g_{2}$ is proportional to both the zero-point fluctuation
$V_{0}$, but (as opposed to the coupling $g_{1}$) also to the amplitude
$\epsilon_{d}$, which allows us to \textit{classically amplify and control}
the spin-resonator interaction strength. Moreover, the fact that (for
a fixed frequency $\omega_{0}$ which does not have to be necessarily
the fundamental mode) the single photon amplitude decreases with the
length of the resonator $L$ as $V_{0}\sim1/\sqrt{L}$ can be compensated
by increasing the driving amplitude $\epsilon_{d}$ (as long as $\delta\epsilon\ll\epsilon_{0}$
to guarantee the validity of the underlying Taylor expansion). 
Following Ref.\cite{harvey18SM}, we have neglected the dispersive shift
term $H_{\mathrm{ds}}=\chi a^{\dagger}a\sigma^{z}$ with $\chi=J''\left(\epsilon_{0}\right)e^{2}c_{r}^{2}V_{0}^{2}/2$,
because (as compared to the longitudinal coupling) this dispersive
coupling is smaller by a factor $\sim ec_{r}V_{0}\bar{n}/\epsilon_{d}\ll1$,
with $\bar{n}=\left\langle a^{\dagger}a\right\rangle $. Depending
on the actual value of $\epsilon_{d}$ the last condtion may impose
a limitation on temperature; however, the effect of the dispersive
shift may also be neglected for sufficiently large detuning in the
limit $\Delta\gg\chi N/2$.


\textit{Multi-Mode Setting.---}As shown in Ref.~\cite{harvey18SM}, the longitudinal spin resonator
coupling can be amplified by parametrically modulating the splitting
with a classical drive. Here, we aim to generalize this idea to a
multi-mode setup. The multi-mode Hamiltonian under consideration reads
\begin{equation}
H=\sum_{n}\omega_{n}a_{n}^{\dagger}a_{n}+\frac{J\left(\epsilon\right)}{2}\sigma^{z},
\end{equation}
with $\epsilon=\epsilon_{0}+\delta\epsilon$, and 
\begin{equation}
\delta\epsilon=\sum_{n}A_{n}\cos\left(\Omega_{n}t\right)+\sum_{n}\phi_{n}\left(a_{n}+a_{n}^{\dagger}\right).
\end{equation}
with $\phi_{n}=ec_{r}V_{n}$ for notational convenience. Here, the
first term describes a polychromatic driving scheme (with amplitudes
$A_{n}$ and frequencies $\Omega_{n}\approx\omega_{n}$) and the second
term describe voltage fluctuations on the dot caused by the multi-mode
resonator. We obtain 
\begin{eqnarray}
\delta\epsilon^{2} & = & \sum_{m,n}A_{m}A_{n}\cos\left(\Omega_{m}t\right)\cos\left(\Omega_{n}t\right)\\
 &  & +\sum_{m,n}\phi_{m}\phi_{n}\left(a_{m}+a_{m}^{\dagger}\right)\left(a_{n}+a_{n}^{\dagger}\right)\\
 &  & +2\sum_{m,n}A_{m}\phi_{n}\cos\left(\Omega_{m}t\right)\left(a_{n}+a_{n}^{\dagger}\right).
\end{eqnarray}
Within the experimentallly most relevant regime we can neglect all
rapidly oscillating terms (provided that $J'\left(\epsilon_{0}\right)A_{n}/4,J'\left(\epsilon_{0}\right)\phi_{n}\sqrt{\left\langle a_{n}^{\dagger}a_{n}\right\rangle }/2\ll\Omega_{n}$,
$J''\left(\epsilon_{0}\right)\phi_{m}\phi_{n}\sqrt{\left\langle a_{m}^{\dagger}a_{m}\right\rangle \left\langle a_{n}^{\dagger}a_{n}\right\rangle }/4\ll\left|\Omega_{m}-\Omega_{n}\right|$
and $J''\left(\epsilon_{0}\right)A_{m}A_{n}/4\ll\left|\Omega_{m}-\Omega_{n}\right|,\Omega_{m}+\Omega_{n}$
$\forall m\neq n$) and keep only co-rotating terms (see the last
line in the expression for $\delta\epsilon^{2}$). In that case, along
the lines of the single-mode analysis the total Hamiltonian simplifies
within a RWA to 
\begin{eqnarray}
H & \approx & \sum_{n}\omega_{n}a_{n}^{\dagger}a_{n}+\frac{\tilde{J}\left(\epsilon_{0}\right)}{2}\sigma^{z}\\
 &  & +\sum_{n}g_{2,n}\left(t\right)\sigma^{z}\otimes\left(a_{n}+a_{n}^{\dagger}\right),
\end{eqnarray}
with 
\begin{equation}
g_{2,n}\left(t\right)=\frac{1}{2}J''\left(\epsilon_{0}\right)ec_{r}V_{n}A_{n}\cos\left(\Omega_{n}t\right).
\end{equation}
This coupling is boosted by the classical amplitude $A_{n}$; in the
lab frame the coupling to mode $n$ oscillates with frequency $\Omega_{n}\approx\omega_{n}$.
In the limit of a single-frequency, monochromatic drive (i.e., $A_{n}=0$
for all but a single mode) we recover the results of the previous
section. In principle we can selectively control the modes the qubit
couples to by choosing the amplitudes $\left|A_{n}\right|\geq0$ appropriately.
For example, by setting $A_{n}=0$ for all even (odd) modes the qubit
couples only to the odd (even) resonator modes. This control can be
done individually for every qubit. 

\subsection{Protocol to implement QAOA}
To realize our scheme for implementing QAOA with this specific setup one could (for example) make use of spin-spin interactions controlled via parametric modulation as detailed in \cite{harvey18SM} together with single-qubit rotations or 
utilize the (tunable) effective electric dipole moment associated with exchange coupled spin states in a DQD \cite{taylor06SM, schuetz17SM}.  
One can then efficiently alternate between the unitaries $U_{x}(\beta)$ and $U_{zz}(\gamma)$ by repeatedly cycling through two parameter regimes:  
(i) The magnetic gradient $\delta B$ dominates over the voltage-controlled qubit frequencies $\omega_{i}(\epsilon) \rightarrow 0$ (with $\epsilon$ denoting the gate voltage), 
resulting in $\beta = \delta B \cdot t_{p}$.
In this regime the coupling to the resonator is turned off \cite{schuetz17SM, harvey18SM}, giving $J_{ij}=0$ and effectively $\gamma_{ij}=0$. 
(ii) Then, by pulsing to a regime where $\delta B$ is the smallest energy scale, we turn on the qubit-qubit coupling $U_{zz}(\gamma_{ij})$ \cite{jin12SM, harvey18SM}. 
This procedure completes one cycle of QAOA.


\begin{thebibliography}{1}
\bibitem{temperature}From a technological point of view, this statement is important 
as thermal occupation of the resonator modes will be inevitable for reasonable temperatures when going to relatively long transmission lines (with correspondingly small fundamental mode frequencies). For example, for a (fundamental) mode frequency of $\sim 100\mathrm{MHz}$ the thermal occupation amounts to $\sim 20$ thermal photons, even at very cold dilution fridge temperatures of $\sim 100\mathrm{mK}$.
\bibitem{eigenvalue}For example, for the two-qubit phase gate there is just a single eigenvalue $w_{q}=\pi/4$, 
such that $t_{p}=t_{\mathrm{coh}}=\pi/4J_{\mathrm{max}}$.
\bibitem{T1decay}We ignore single spin relaxation processes, since the associated timescale $T_{1}$ is typically much longer than $T_{2}$. For example, for single-electron spins in silicon $T_{1}$ of up to $\sim 3\mathrm{s}$ has been demonstrated \cite{simmons11},  while $T_{2} \lesssim 10 \mathrm{ms}$ \cite{veldhorst15}.  Still, if necessary, $T_{1}$ processes could be included along the lines of dephasing-induced errors \cite{schuetz17}.
\end{thebibliography}
\end{document}